\documentclass[pdflatex,sn-mathphys-num]{sn-jnl}

\usepackage{graphicx}%
\usepackage{multirow}%
\usepackage{amsmath,amssymb,amsfonts}%
\usepackage{amsthm}%
\usepackage{mathrsfs}%
\usepackage[title]{appendix}%
\usepackage{xcolor}%
\usepackage{textcomp}%
\usepackage{manyfoot}%
\usepackage{booktabs}%
\usepackage{algorithm}%
\usepackage{algorithmicx}%
\usepackage{algpseudocode}%
\usepackage{listings}%
\usepackage{acronym}
\usepackage{verbatim}

\usepackage{array}  
\usepackage{colortbl} 
\usepackage{rotating}
\usepackage{adjustbox}
\usepackage{hhline} 

\usepackage{longtable}
\usepackage{pdflscape}
\usepackage{lscape}
\usepackage[hyphenbreaks]{breakurl}

\theoremstyle{thmstyleone}%
\theoremstyle{thmstyletwo}%

\theoremstyle{thmstylethree}%

\begin{document}

\title[Article Title]{Identification of Malicious Posts on the Dark Web Using Supervised Machine Learning}


\author[1]{\fnm{Sebastião} \sur{Alves de Jesus Filho}}\email{sebastiao@ufu.br}

\author[1]{\fnm{Gustavo} \sur{Di Giovanni Bernardo}}\email{gustavo.bernardo@ufu.br}

\author[1]{\fnm{Paulo} \sur{Henrique Ribeiro Gabriel}}\email{phrg@ufu.br}

\author[2]{\fnm{Bruno} \sur{Bogaz Zarpelão}}\email{brunozarpelao@uel.br}

\author*[1]{\fnm{Rodrigo} \sur{Sanches Miani}}\email{miani@ufu.br}



\affil*[1]{\orgdiv{Faculty of Computing}, \orgname{Universidade Federal de Uberlândia}, \orgaddress{\street{Av. João Naves de Ávila, 2121 - Santa Mônica}, \city{Uberlândia}, \postcode{38408-100}, \state{Minas Gerais}, \country{Brasil}}}

\affil[2]{\orgdiv{Computer Science Department}, \orgname{State University of Londrina}, \orgaddress{\street{Rod. Celso Garcia Cid s/n}, \city{Londrina}, \postcode{86057-970}, \state{Parana}, \country{Brazil}}}




\abstract{

Given the constant growth and increasing sophistication of cyberattacks, cybersecurity can no longer rely solely on traditional defense techniques and tools. Proactive detection of cyber threats has become essential to help security teams identify potential risks and implement effective mitigation measures. Cyber Threat Intelligence (CTI) plays a key role by providing security analysts with evidence-based knowledge about cyber threats. CTI information can be extracted using various techniques and data sources; however, machine learning has proven promising. As for data sources, social networks and online discussion forums are commonly explored.
In this study, we apply text mining techniques and machine learning to data collected from Dark Web forums in Brazilian Portuguese to identify malicious posts. Our contributions include the creation of three original datasets, a novel multi-stage labeling process combining indicators of compromise (IoCs), contextual keywords, and manual analysis, and a comprehensive evaluation of text representations and classifiers. To our knowledge, this is the first study to focus specifically on Brazilian Portuguese content in this domain. The best-performing model -- using LightGBM and TF-IDF -- was able to detect relevant posts with high accuracy. We also applied topic modeling to validate the model’s outputs on unlabeled data, confirming its robustness in real-world scenarios.

}




\maketitle

\section{Introduction}\label{sec_introduction}

Illicit activities on the Internet, such as financial data theft, extortion, and cyber espionage, have significantly increased \cite{Brooks2021-cq}. Serious incidents, such as the ransomware attacks against Colonial Pipeline \cite{Satter_Reuters_Colonial}, ICBC \cite{Schroeder_Reuters_ICBC}, and Synnovis \cite{Mercer2024-zk_Synnovis}, the RockYou password leak \cite{securityleaders_RockYou2024}. The exposure of Brazilian citizens' data \cite{Marin2021-ty_vazamento_br} clearly illustrates cyberattacks' financial and social impact. The rise in cyberattacks in recent years has resulted partly from the COVID-19 pandemic, as many organizations adopted remote work without implementing the necessary security measures against these attacks \cite{dutta2020overview}. Hacker communities have reported a growing number of posts discussing the exploitation of the pandemic as a new opportunity for attacks, mainly targeting remote work tools \cite{mador2021keep}.

The Dark Web, known for preserving anonymity, has become a conducive environment for exchanging information among cybercriminals, facilitating coordinated cyberattacks and other illegal activities, such as data breaches and ransomware attacks \cite{saleem2022anonymity}. With the advancement of hacking techniques and tools, these actions have shifted from isolated acts to organized, often funded, operations with financial or political motivations \cite{tounsi2019cyber}. Experts warn of the urgent need for effective preventive measures to protect organizations from these increasing attacks \cite{basheer2021threats}.

Traditional tools such as packet filters and intrusion detection systems no longer seem sufficient to prevent information compromise in this scenario. The advancement of computational power in digital systems, combined with the improved Tactics, Techniques, and Procedures (TTPs) employed by cybercriminals, has rendered conventional security controls insufficient for detecting intrusions and preventing threats in the current cybersecurity landscape \cite{dutta2020overview}. As a result, more researchers and cybersecurity professionals are focusing on a new generation of on-demand cybersecurity tools known as Cyber Threat Intelligence (CTI) \cite{basheer2021threats}. CTI collects data to provide evidence-based knowledge about threats, helping organizations detect, prevent, and recover from cyberattacks \cite{tounsi2019cyber}, \cite{sari2018context}.

According to Sapienza et al. \cite{sapienza2017early}, malicious actors follow a series of steps to conduct cyberattacks. These steps include identifying vulnerabilities, acquiring tools and skills, selecting a target, creating or obtaining infrastructure, and planning and executing the attack. During these phases, malicious actors may leave traces associated with specific activities, such as attempts to access unusual URLs or manipulate corporate email lists. These traces are known as Indicators of Compromise (IoCs) \cite{jo2022vulcan}, functioning as a sort of fingerprint that information security experts can observe. Further examples of IoCs include Internet Protocol (IP) addresses, domain names, and file hashes.

IoCs are essential for CTI sharing, but they should not be the sole focus, as they can become outdated over time \cite{jo2022vulcan}, \cite{preuveneers2021sharing}. With the rise in cyber threats, enhancing CTI sources, including the analysis of Dark Web content, is necessary \cite{basheer2021threats}. Additionally, integrating advanced technologies such as machine learning is crucial to processing large volumes of data and enabling quick and proactive responses to cyber threats \cite{preuveneers2021sharing}, \cite{basheer2021threats}.

Between 2017 and 2023, various studies investigated security threats using different sources and methods. Some focused on analyzing content from Dark Web forums~\cite{sapienza2017early},~\cite{dong2018new},~\cite{sarkar2019predicting}, and~\cite{arnold2019dark}. In the context of IoC analysis and extraction, Niakanlahiji et al.~\cite{niakanlahiji2019iocminer} investigated the presence of IoCs on Twitter, while Zhang et al.~\cite{zhang2019imcircle} extracted IoCs from the Surface Web, using indicators such as domains and IP addresses considered suspicious based on open-source threat intelligence. Meanwhile, Caballero et al.~\cite{caballero2023rise} extracted IoCs from six different sources, including blogs, Twitter, Telegram, and structured repositories such as Malpedia, APTnotes, and ChainSmith. However, these works did not conduct IoC studies using Dark Web data.

Other studies, such as those by Queiroz et al.~\cite{queiroz2019detecting} and Koloveas et al.~\cite{koloveas2021intime}, applied machine learning to content collected from the Dark Web to derive CTI. Despite their contributions, those studies left open challenges regarding the identification of relevant content and, notably, the lack of publicly available labeled datasets. In addition, their labeling methodologies were either insufficiently described or based solely on keyword matching, without integrating IoCs and manual expert validation.

In this context, Brazilian Portuguese was selected as the target language due to its growing relevance in cybercriminal ecosystems. Recent threat intelligence reports show that Brazil ranks among the most frequently cited countries in Dark Web marketplaces and ransomware incidents across Latin America~\cite{socRadarbrazil2024},~\cite{kasperskyBrazil2025}. Yet, despite the country’s exposure, CTI datasets and classification tools in Portuguese remain scarce. Addressing this gap contributes to the global advancement of multilingual CTI research and supports the creation of more inclusive detection and automation mechanisms.

Moreover, the unavailability of labeled data remains a challenge. Although Queiroz et al.~\cite{queiroz2019detecting} provided a URL for dataset download, it was no longer active during the development of this study. Additionally, those works did not evaluate modern gradient boosting classifiers such as LightGBM~\cite{ke2017lightgbm} and XGBoost~\cite{chen2016xgboost}, nor did they test models using unlabeled data to assess generalization capabilities.

Given the challenges discussed earlier, this work aims to contribute with the following:

\begin{enumerate}

\item Development of a tool for extracting IoCs from unstructured data sources, such as the Dark Web;

\item Creation of a Brazilian Portuguese labeled dataset for training supervised machine learning models in the context of cybersecurity;

\item Development of a classification model for identifying relevant posts to the cybersecurity community.

\end{enumerate}
 
The remainder of the article is structured as follows: Section \ref{sec_background} presents the concepts of CTI, the Deep Web, the Dark Web, and Text Classification. Section \ref{sec_related_work} reviews relevant works that contributed to the development of this research. Section \ref{sec_materiais_metodos} describes the method used to identify malicious posts on the Dark Web by applying supervised machine learning techniques. Section \ref{sec_resultados} discusses the experiments conducted and the results obtained, while Section \ref{sec_conclusao} presents the article's conclusions.

\section{Background}\label{sec_background}

\subsection{Cyber Threat Intelligence} \label{subsec_cti}

Cyber Threat Intelligence (CTI) is an evidence-based approach that enables organizations to proactively identify, understand, and mitigate cyber threats~\cite{saxena2022cyber}. It involves the continuous monitoring and analysis of information sources, such as news sites, social media, blogs, and Dark Web forums, to transform raw data into actionable knowledge~\cite{jo2022vulcan}. A key element of CTI is the use of Indicators of Compromise (IoCs), which function as digital fingerprints -- examples include IP addresses, domain names, file hashes, and URLs~\cite{sun2023cyber}.

IoCs are commonly classified into three categories~\cite{niakanlahiji2019iocminer}: \textit{atomic} (e.g., IP addresses), \textit{computed} (e.g., malware hash values), and \textit{behavioral}, which combine indicators of activity and context. Understanding and correlating these elements help reduce incident response times and improve detection~\cite{asiri2023understanding}. However, as highlighted by Jo et al.~\cite{jo2022vulcan}, IoCs alone are insufficient to capture the dynamic and evolving nature of cyber threats. Therefore, CTI should be part of a broader strategy that incorporates contextual analysis and adaptive methodologies.

\subsection{Deep Web and Dark Web} \label{subsection_deepweb_darkweb}

The Surface Web refers to the publicly accessible portion of the Internet indexed by common search engines such as Google and Yahoo, and accessible through conventional browsers like Mozilla, Edge, and Opera. According to Akhgar et al.~\cite{akhgar2021dark}, this layer has existed since the first graphical browsers were developed. In contrast, the Deep Web includes content that is not indexed by search engines, such as password-protected banking portals and private email services. Although publicly available, these resources require authentication or encryption to access. The Dark Web represents the most restricted portion of the Deep Web, where both users and servers rely on anonymization technologies (such as Tor) to conceal identities and locations, making this content unreachable through standard access methods.

While anonymity on the Dark Web serves legitimate purposes, such as bypassing censorship in authoritarian regimes, it has also enabled a wide range of illicit activities~\cite{bradbury2014unveiling,basheer2021threats}. These include marketplaces for illegal goods and services, forums for hacker collaboration, and platforms for financial crimes. Saleem et al.~\cite{saleem2022anonymity} emphasize that the technical structure of the Dark Web facilitates this underground economy, offering attackers a high degree of concealment. In this work, we focus specifically on open-access Dark Web forums that do not require credentials, which are commonly used to share threat-related information and services.

\subsection{Dark Web Forum} \label{subsec_forum_darkweb}

A Dark Web forum is an online platform where users discuss various topics, typically divided into categories, each dedicated to specific themes such as hacking, drugs, money, and markets. However, according to Al-Ramahi et al. \cite{al2020exploring}, users do not always adhere to these predefined categories and may post content related to a specific topic in a different category. Akhgar et al. \cite{akhgar2021dark} emphasize that accessing a forum on the Dark Web requires specialized browsers, such as The Onion Router (Tor), which routes Internet traffic through multiple servers worldwide, making it difficult to trace the user's identity.

To create content in a Dark Web forum, users follow basic steps: they select a category, define a title, and, in a separate field, detail the content of their post. Both the title and the content often include shared information or questions. The initial post can receive responses and comments from other users, known as interactions, depending on the relevance of the content. Although any response, comment, or initial message can be considered a post, in this work, we define a post as a set of messages related to a specific topic. The number of interactions or whether they occur does not affect the definition of a post.

In this work, we consider a post \emph{malicious} or \emph{relevant} if its content relates to any threat, vulnerability, exploit, data leak, or anything that indicates a risk to cybersecurity. Otherwise, we classify it as \emph{not relevant}.

\subsection{Text Classification} \label{subsec_text_class}

Texts collected from Dark Web forums form unstructured datasets that require preprocessing and transformation before they can be used for classification. Text mining techniques~\cite{dang2014text} are commonly applied to extract meaningful patterns from this type of data. A key step in this process is the conversion of text into numerical representations suitable for machine learning algorithms.

Among the most common techniques is Term Frequency--Inverse Document Frequency (TF-IDF), which quantifies the relevance of terms by balancing their frequency in individual documents and across the corpus~\cite{qaiser2018text}. Alternatives such as word2vec~\cite{mikolov2013efficient} have also been employed to capture semantic relationships between words through dense vector embeddings. In this study, we adopt TF-IDF for feature extraction, due to its interpretability and effectiveness in high-dimensional text classification tasks.

\section{Related Work}\label{sec_related_work}

This section presents the main works related to this article, aiming to provide an overview of the state of the art in information extraction for CTI using unstructured data sources, such as social networks and Dark Web forums. These proposals work from the assumption that it is possible to identify signs of existing threats from these sources and even predict new cyberattacks before they occur.

Some works have explored social networks and the Dark Web to identify cyber threats, employing various approaches. Others have focused on information extraction for CTI through IoCs, while some have used machine learning to extract CTI from the Dark Web. This study focused on identifying malicious posts using text mining techniques and supervised machine learning on data collected from Dark Web forums. This process shares some common steps with related works, as they all aim to obtain CTI from unstructured data sources, such as social networks, Surface Web, and Dark Web forums. Table \ref{tab:comparative_related_work} summarizes the related work.

\begin{sidewaystable}[htbp]
\centering
\footnotesize
\caption{Comparison between related works and the proposed study}
\begin{tabular}{|l|p{2.8cm}|l|c|c|p{4.8cm}|}
\hline
\textbf{Reference} & \textbf{Data Source(s)} & \textbf{Language(s)} & \textbf{Public Dataset?} & \textbf{ML Used?} & \textbf{Main Objective} \\
\hline
\cite{sapienza2017early} & Twitter + Dark Web forums & English & No & Partial & Early cyber threat alerts from expert tweets \\
\hline
\cite{alves2021processing} & Twitter & English & No & Yes & Infrastructure-focused threat summarization \\
\hline
\cite{rodriguez2020social} & Twitter & English & No & Yes & Real-time alerting via sentiment classification \\
\hline
\cite{arnold2019dark} & Dark Web forums & EN, RU, FI & No & No & Identify threats using social network graphs \\
\hline
\cite{niakanlahiji2019iocminer} & Twitter & English & No & Yes & Automatic IoC extraction using user reputation \\
\hline
\cite{zhang2019imcircle} & Surface Web & English & No & Yes & Automated verification of IoCs from the web \\
\hline
\cite{al2020exploring} & Dark Web & English & Yes (public dataset used) & Partial & Extract Topics of Interest (ToIs) from forums \\
\hline
\cite{caballero2023rise} & Blogs, Telegram, Twitter, Malpedia & English & No & Yes & IoC extractor development and comparison \\
\hline
\cite{sarkar2019predicting} & Dark Web & English & No & Yes & Predict enterprise attacks using CVE graphs \\
\hline
\cite{queiroz2019detecting} & Dark + Surface Web & English & URL inactive & Yes & Compare embeddings for malicious post detection \\
\hline
\cite{koloveas2021intime} & Dark + Surface Web & English & No & Yes & Rule-based CTI extraction focused on IoT \\
\hline

\hline
\cite{alevizos2024ai} & Open CTI sources (multi-stage pipeline) & English & No & Yes & AI-powered automation of CTI ingestion, analysis and recommendation \\

\hline
\cite{bala2025unveiling} & Dark Web (multi-language forums) & Multiple & No & Yes & Enhance ML models to decode illicit communication in the Dark Web \\

\hline

\textbf{This Work} & Dark Web forums (Hidden Answers, Deep Answer) & \textbf{Brazilian Portuguese} & \textbf{Yes} & \textbf{Yes} & \textbf{Identify malicious posts using labeled data and classify new content} \\
\hline
\end{tabular}
\label{tab:comparative_related_work}
\end{sidewaystable}


Sapienza et al. \cite{sapienza2017early} proposed a framework that leverages social media sensors, particularly Twitter and Dark Web forums, to generate early warnings of cyber threats. The authors employed a strategy of monitoring the Twitter accounts of manually selected experts, researchers, and ethical hackers to find posts related to vulnerability exploitation. Using text mining techniques, they selected important terms and removed irrelevant ones based on predefined dictionaries. They then checked if the discovered terms appeared in previously selected Dark Web hacker forums. During the observation period, experts considered 84\% of the generated warnings relevant. The authors reported significant security events, such as the Mirai attack in October 2016, which exploited the vulnerabilities of Internet of Things (IoT) devices, and data breaches like AdultFriendFinder and BrazzersForum, which emerged during the tests.

In a similar way, Alves et al. \cite{alves2021processing} presented a threat monitor that uses Twitter to generate a continuously updated summary of the threat landscape related to a monitored infrastructure. Initially, the system works with a manually created list of keywords describing the information technology (IT) infrastructure and a predefined list of Twitter accounts related to cybersecurity. The system uses an Application Programming Interface (API) to capture posts from the monitored accounts and applies a filter based on the keyword list. Only tweets with at least one word from the list pass through the filter. The system then prepares the data for processing using text-mining techniques. Next, Supervised machine learning algorithms classify the tweets according to their security relevance. Before generating security alerts, a clustering stage groups similar tweets and retweets. The process revealed a challenge in database labeling for training, as an analyst had to mark all irrelevant tweets manually.

Meanwhile Rodriguez and Okamura \cite{rodriguez2020social} present a real-time system that uses data analysis from Twitter to aggregate large amounts of tweets and generate cybersecurity awareness information. The system analyzes tweet context through sentiment analysis to assess the threat risk level. The authors selected cybersecurity-related Twitter accounts during data collection and applied a keyword filter. The system compares tweet content against a security keyword list to ensure contextual relevance. TF and TF-IDF techniques help discover new keywords. Based on supervised machine learning, a sentiment classifier categorizes tweets as negative or positive. Analysts can monitor the current situation through a detailed graphical interface, where a high volume of negative tweets containing specific keywords might signal an ongoing attack.

Furthermore Arnold et al. \cite{arnold2019dark} proposed a CTI tool that uses graph analysis and complex networks to identify cyber threats from Dark Web data sources. The authors reported using data collected from eight English, Russian, and Finnish forums to form a multi-node social network. Through SQL queries, they identified 132 organization names mentioned in these forums, including well-known companies like \emph{Amazon, PayPal}, and \emph{Microsoft}. The authors reported identifying a large number of threats, with the highest number related to fraud, followed by account breaches and hacking tools available for attacks on companies and their customers.


Niakanlahiji et al. \cite{niakanlahiji2019iocminer} presented a scalable framework for automatic IoC extraction from Twitter, using a combination of graph theory, machine learning, and text mining techniques. The system includes a reputation model to discover reliable profiles that publish CTI information and only tracks the tweet flow from those profiles. The authors reported that, over four weeks, the system identified more than 1,200 IoCs, including malicious URLs.

In a related approach, Zhang et al. \cite{zhang2019imcircle} presented a system capable of automatically extracting IoCs from the Surface Web, verifying suspicious indicators with the help of open-source threat intelligence. The system receives suspicious indicators, such as domains and IP addresses, and checks if they are related to actual threats by actively collecting and analyzing their relevant threat information from the Surface Web. Based on the verification results, the system generates a list of IoCs. It then automatically extracts new indicators from web pages related to the IoCs as new inputs, repeating the verification process to generate more IoCs.

In another study Al-Ramahi et al. \cite{al2020exploring} presented a systematic approach to automatically extract Topics of Interest (ToIs) from hacker websites, aiming to use them as inputs for actionable security controls or IoC collectors. Initially, the authors analyzed hacker posts in a public dataset. They developed a tracker to extract ToIs from a Dark Web forum as a second experiment. The results were positive; however, the authors reported several challenges related to tracking and extracting relevant ToIs.

Moreover Caballero et al. \cite{caballero2023rise} introduced a platform to extract IoCs from six different sources: \emph{Blogs RSS, Twitter, and Telegram}, as well as \emph{Malpedia, APTnotes}, and \emph{ChainSmith}, which are repositories for cybersecurity-related projects. In addition to developing the IoC extraction tool, the authors also analyzed the accuracy of seven other IoC extraction tools. The results showed their developed tool achieved higher accuracy in 11 of the 13 types of IoCs extracted.


Sarkar et al. \cite{sarkar2019predicting} have used data from Dark Web forums, analyzing the structure of user responses to predict corporate cyberattacks. This structure captures how interactions are connected, forming a network or graph. The proposed system attempts to predict whether a cyberattack will occur on a given day for an organization by applying supervised learning models to a set of features extracted from the forums. The authors acquired the Dark Web data through a commercial programming interface. First, they selected a set of forums they considered most relevant. Then, they searched for mentions of vulnerabilities in this set of forums, thus computing the total number of CVEs mentioned in these posts. They grouped the CVEs using the structured nomenclature scheme of Common Platform Enumeration (CPE) from the NVD database maintained by the National Institute of Standards and Technology (NIST). The system uses directed graphs to extract a set of specialized users, called \emph{experts}, whose posts mentioning vulnerabilities capture the attention of other users over a specific period. The system then generates a time series to capture the interactions of these expert users in Dark Web forums. Finally, a learning model attempts to predict cyberattacks before they occur.

Following a similar direction Queiroz et al. \cite{queiroz2019detecting} proposed an approach to improve classification models using language models for feature representation, employing word and sentence embedding techniques to identify contextual semantic properties in words and phrases, enabling the detection of cyber threats related to vulnerabilities in forums and social networks on the Surface Web and Dark Web. The authors aimed to investigate the performance of embedding models, such as Word2Vec, in detecting hacker threats in online forums, comparing them to classical language models. They tested combinations of pre-trained models Word2Vec, GloVe, Sent2Vec, InferSent, and SentEncoder together with two supervised learning algorithms, SVM and Convolutional Neural Networks (CNN), using data from five sources, including four Dark Web forums and Twitter, totaling 9,470 samples. After manual labeling, they classified around 11.8\% of the messages as malicious. Due to the low recall rate in the initial tests, class-balancing techniques raised this rate to 37.2\%. The Word2Vec-based model achieved the best performance, reaching 96\% accuracy and 93\% recall, surpassing previous results.

Similarly Koloveas et al. \cite{koloveas2021intime} have presented an integrated framework for mining and extracting Cyber Threat Intelligence (CTI) from various sources, including the Surface Web, social networks, and the Dark Web. The framework performs four primary tasks: collecting, analyzing, managing, and sharing CTI data. The authors implemented Word2Vec for text representation along with two supervised machine-learning algorithms. They created a labeled dataset focusing on Internet of Things (IoT) security for model training. They established a simple rule for page classification: pages containing both \emph{Security} and \emph{IoT} terms qualified as relevant, while those containing only one term qualified as irrelevant.

The approach in this article shares some similarities with related works, from those that aimed to identify cyber threats using different approaches where the primary data source was Twitter to those that focused on CTI extraction through IoCs, and especially those that employed supervised machine learning and used the Dark Web. However, our proposal encompasses these sub-areas while introducing significant differences. Notable aspects include our specific choice of forums in Brazilian Portuguese, our data labeling process involving the identification of IoCs, contextual keywords, and manual analysis. We make the labeled dataset available and conduct tests with different machine-learning algorithms using various forms of text representation. This work stands out mainly due to our testing methods and analysis of how the system classifies new unlabeled posts.

While utilizing the Dark Web as a data source, the approaches proposed by \cite{sapienza2017early} and \cite{arnold2019dark} did not employ machine learning for threat identification. Conversely, the studies by \cite{alves2021processing} and \cite{rodriguez2020social} are notable for selecting Twitter as the primary data source for identifying cyber threats. In the context of information extraction for Cyber Threat Intelligence (CTI) using Indicators of Compromise (IoCs), the studies by \cite{niakanlahiji2019iocminer}, \cite{zhang2019imcircle}, \cite{al2020exploring}, and \cite{caballero2023rise} did not specifically focus on extracting IoCs from Dark Web forums, as undertaken in this article during the data labeling phase.

More recent works have explored the integration of artificial intelligence into the CTI lifecycle. Alevizos et al. \cite{alevizos2024ai} proposed an AI-enhanced pipeline that automates CTI ingestion, analysis, and recommendation steps, aiming to improve both speed and scalability. Shah and Khoda Parast \cite{shah2024ai} investigated the use of LLMs fine-tuned in a one-shot fashion to automate CTI generation for industrial environments, reducing human effort while maintaining high analytical accuracy. Ali et al. \cite{ali2025privacy} developed a privacy-aware framework based on federated learning and graph neural networks to analyze cyberterrorism networks without centralizing sensitive data. Meanwhile, Bala et al. \cite{bala2025unveiling} applied machine learning to decode illicit communication patterns in multilingual Dark Web forums. These studies reflect the growing role of advanced AI techniques in threat intelligence.

The studies by \cite{sarkar2019predicting}, \cite{queiroz2019detecting},  \cite{koloveas2021intime}, and \cite{bala2025unveiling} closely align with this article, as they similarly leverage machine learning techniques and utilize the Dark Web as a primary data source. However, none address the challenges of analyzing Brazilian Portuguese content or describe a labeling strategy that integrates IoCs, contextual keywords, and manual validation as proposed in our work.

\section{Materials and Methods}\label{sec_materiais_metodos}

This section details our method to identify malicious posts on the Dark Web by applying supervised machine learning techniques. We divided the work into three main development phases. The first phase involved constructing labeled datasets, the second involved developing a post classification model, and the third included testing the model on a new unlabeled dataset. Subsection \ref{subsection_construcao_datasets_rot} describes how we constructed the labeled datasets, while Subsection \ref{subsection_desenvol_mod_classificacao} discusses the development and testing of the post classification model. All datasets generated and analyzed during this study (``DarkPT-BR: Labeled Posts from Brazilian Portuguese Dark Web Forums'') are publicly available at Mendeley Data \cite{jesus2025darkptbr}.

\subsection{Construction of Labeled Datasets}\label{subsection_construcao_datasets_rot}

The first phase of this work focused on constructing labeled datasets to train supervised machine learning algorithms. This step became necessary due to the scarcity of labeled data and the specific project challenges, similar to works such as \cite{dong2018new}, \cite{koloveas2021intime}, and \cite{queiroz2019detecting}, which also created their own labeled datasets. Although \cite{queiroz2019detecting} provided a URL for their labeled dataset, we could not access it during this work's development. Moreover, even with access to this data, it would not suit this study's requirements, as we chose to analyze posts from forums in Brazilian Portuguese.

This phase of the work begins with the collection of posts, followed by initial pre-processing, IoC extraction, second pre-processing, and topic modeling, leading to the data labeling process, as described in the following subsections.

\subsubsection{Stage I - Post Collection} \label{subsubsection_coleta}

The first stage of the process involves collecting posts from two Brazilian Portuguese-language Dark Web forums: \emph{Hidden Answers} and \emph{Deep Answer}. We selected these forums for their open nature, as any user with the corresponding link or URL can access them. Unlike restricted or private forums, which require invitations or passwords for access, open forums provide easier accessibility and tend to have a higher flow of information and activity.

To collect the data, we used a crawler, i.e., an automated system that scans the forums for posts. The system stores all the collected data in a database in JSON format, with the attributes listed in Table \ref{tab_atributos_foruns}. For this study, the system collected 26,575 posts.

\begin{table}[!ht]
\centering
\caption{Attributes present in the JSON files collected from the \textit{Hidden Answers} and \textit{Deep Answers} forums}
\label{tab_atributos_foruns}
\begin{tabular}{r l @{\hskip 2cm} r l}
\hline
\multicolumn{2}{c}{\textbf{Forum \textit{Hidden Answers}}} & \multicolumn{2}{c}{\textbf{Forum \textit{Deep Answers}}} \\ \hline
\multicolumn{1}{r}{1} & \multicolumn{1}{l}{category} & \multicolumn{1}{r}{1} & \multicolumn{1}{l}{category} \\ \hline
\multicolumn{1}{r}{2} & \multicolumn{1}{l}{title} & \multicolumn{1}{r}{2} & \multicolumn{1}{l}{title} \\ \hline
\multicolumn{1}{r}{3} & \multicolumn{1}{l}{content} & \multicolumn{1}{r}{3} & \multicolumn{1}{l}{question} \\ \hline
\multicolumn{1}{r}{4} & \multicolumn{1}{l}{answers} & \multicolumn{1}{r}{4} & \multicolumn{1}{l}{answers} \\ \hline
\multicolumn{1}{r}{5} & \multicolumn{1}{l}{created\_at} & \multicolumn{1}{r}{5} & \multicolumn{1}{l}{dataCreated} \\ \hline
\multicolumn{1}{r}{6} & \multicolumn{1}{l}{author} & \multicolumn{1}{r}{6} & \multicolumn{1}{l}{author} \\ \hline
\multicolumn{1}{r}{7} & \multicolumn{1}{l}{tags} & \multicolumn{1}{r}{7} & \multicolumn{1}{l}{tags} \\ \hline
\multicolumn{1}{r}{8} & \multicolumn{1}{l}{comments} & \multicolumn{1}{r}{8} & \multicolumn{1}{l}{type} \\ \hline
\multicolumn{1}{r}{9} & \multicolumn{1}{l}{best\_answer} & \multicolumn{1}{r}{9} & \multicolumn{1}{l}{votes} \\ \hline
\multicolumn{1}{r}{10} & \multicolumn{1}{l}{up\_votes} & \multicolumn{1}{r}{10} & \multicolumn{1}{l}{points} \\ \hline
\multicolumn{1}{r}{11} & \multicolumn{1}{l}{down\_votes} & \multicolumn{1}{r}{} & \multicolumn{1}{l}{} \\ \hline
\end{tabular}
\end{table}

It is important to note that data used in this study were collected exclusively from publicly accessible Dark Web forums that do not require user authentication or circumvention of access controls. The crawling process was limited to textual content publicly displayed on open threads. No attempts were made to de-anonymize users or collect personal identifiers.
The research complies with Brazilian data protection principles and does not involve the storage or redistribution of sensitive or illegal content. Manual analysis was conducted with caution to avoid exposure to harmful material. Our aim is to support cybersecurity research while respecting the privacy and anonymity of individuals in these environments.

\subsubsection{Stage II - Initial Pre-processing} \label{subsubsection_pre_proc_I}

This pre-processing stage, implemented using the \emph{pandas} library in \emph{Python}, aims to prepare the data for IoC extraction (Subsection \ref{subsubsection_extra_iocs}). As shown in Table \ref{tab_atributos_foruns}, the JSON files from each forum have different structures -- for instance, attribute 3 in both \emph{Hidden Answers} and \emph{Deep Answers} forums contain equivalent information but uses different names (\emph{content} in one forum and \emph{question} in the other). Meanwhile, attribute 5, which stores the post creation date, differs in name and format across forums. In this stage, the data is grouped into a single file, performing attribute name standardization, creating a new attribute called \emph{full\_text} through concatenating text fields, and adding sequential ID numbers for each post. Table \ref{tab_atributos_definidos} shows the unified attributes after this pre-processing stage.

\begin{table}[!ht]
\centering
\caption{Defined attributes in the pre-processing stage}
\label{tab_atributos_definidos}
\begin{tabular}{lll|l}
\hline
\multicolumn{3}{c|}{\textbf{Attribute}}                             & \multicolumn{1}{c}{\textbf{Description}}                          \\ \hhline{|---|-|}
\multicolumn{1}{l|}{1} & \multicolumn{2}{l|}{ID}                   & Sequential code of the posts            \\ \hhline{|---|-|}
\multicolumn{1}{l|}{2} & \multicolumn{2}{l|}{category}             & Category in which the post was included   \\ \hhline{|---|-|}
\multicolumn{1}{l|}{}  & \multicolumn{1}{l|}{}                     & title   & Contains the post title                \\ \hhline{|~~|-|-|}
\multicolumn{1}{l|}{3} & \multicolumn{1}{l|}{full\_text}           & content & Contains the main text of the post      \\ \hhline{|~~|-|-|}
\multicolumn{1}{l|}{}  & \multicolumn{1}{l|}{}                     & answers & Contains user replies and comments     \\ \hhline{|---|-|}
\multicolumn{1}{l|}{4} & \multicolumn{2}{l|}{created\_at}          & Contains the date when the post was created \\ \hline
\end{tabular}
\end{table}

\subsubsection{Stage III - IoC Extraction} \label{subsubsection_extra_iocs}

The third phase of this work involves the development of the IoC extraction module using the \emph{pandas}\footnote{Official website: \url{https://pandas.pydata.org/}} and \emph{re}\footnote{Documentation: \url{https://docs.python.org/3/library/re.html}} libraries in \emph{Python}. The primary purpose of this module is to identify and flag all posts containing IoCs, with this flagging as one of the parameters used for subsequent data labeling. Our analysis uses regular expressions tailored to each type of IoC to identify patterns in the data. For instance, a specific regular expression identifies IoCs of the email type:
\begin{center}
\verb|(r"[a-zA-Z0-9.]+@[a-zA-Z0-9]+.[a-zA-Z]+(.[a-zA-Z]+)*")|
\end{center}

We extracted a portion of the IoCs using \emph{ioc-finder}\footnote{ioc-finder: \url{https://github.com/fhightower/ioc-finder}}, version 7.2.4, an open-source tool that Forrest Hightower \cite{ioc-finder} developed and made available on \emph{GitHub}. Integrating this tool into our project enhanced our capability to search and extract various types of indicators, improving our threat analysis process. This work focuses on atomic-type IoCs, including IP addresses, emails, URLs, and domain names.

Table \ref{tab_tipos_iocs_procurados} shows the types of IoCs we searched for in each post and the tool we used for searching. The system creates an attribute for each type of IoC defined in Table \ref{tab_tipos_iocs_procurados} to track its presence or absence in each post. Additionally, it creates another attribute named IOC to indicate whether it finds at least one IoC in the post. The system initially sets all values of this attribute to \emph{NO}.

\begin{table}[!ht]
\centering
\caption{Types of IoCs searched and the extraction tool used}
\label{tab_tipos_iocs_procurados}
\begin{tabular}{|lll}
\hline
\multicolumn{2}{l|}{\textbf{Type of IoC}}                                           & \textbf{Search Tool} \\ \hline
\multicolumn{2}{l|}{URL}                                                            & Own Tool                      \\ \hline
\multicolumn{2}{l|}{E-mail}                                                         & Own Tool                      \\ \hline
\multicolumn{2}{l|}{Domain}                                                        & Own Tool                      \\ \hline
\multicolumn{1}{l|}{\textit{Hash}} & \multicolumn{1}{l|}{MD5, SHA1, SHA256, SHA512 e SSDEEP} & IoC-Finder                   \\ \hline
\multicolumn{2}{l|}{IPv4}                                                           & Own Tool                      \\ \hline
\multicolumn{2}{l|}{IPv6}                                                           & IoC-Finder                   \\ \hline
\multicolumn{2}{l|}{ASN}                                                            & IoC-Finder                   \\ \hline
\multicolumn{2}{l|}{CVE}                                                            & Own Tool                      \\ \hline
\multicolumn{2}{l|}{MAC}                                                            & IoC-Finder                   \\ \hline
\multicolumn{2}{l|}{\textit{Registry Key Path}}                                              & IoC-Finder                   \\ \hline
\end{tabular}
\end{table}

Next, the system scans each post, and when it finds at least one IoC, it marks the corresponding column with the value \emph{1} and updates the \emph{IOC} column to \emph{YES}. As previously mentioned, we will use this marking as one of the parameters for labeling the database to train supervised machine learning models.

At the end of the process, we performed a manual check to eliminate IoCs that had similar formats but did not correspond to legitimate IoCs. For example, while our system identified the sequence \emph{4.2.0.2} during the search due to its format being compatible with an IPv4 address, in this specific case, it referred to a software version.

\subsubsection{Stage IV - Preprocessing for Text Mining} \label{subsubsection_pre_proc_II}

As described in Subsection \ref{subsubsection_pre_proc_I}, the data preparation for IoC extraction differs from the data preparation for machine learning models. Text mining considers certain characters undesirable, such as dots and at signs, even though these characters are essential for specific IoCs like IPv4 addresses and emails. This requirement led us to implement two distinct preprocessing stages in our study.

In this stage, we cleaned the data by removing several elements: special characters, numbers, irrelevant terms (such as \emph{QuestionID} and \emph{AnswerID}), and repetitive sequences (like \emph{kkkkkk} and \emph{aaaaaaa}). We also eliminated stopwords using the stopwords package from the \emph{nltk} library, along with HTML tags, URLs, and additional whitespace. Additionally, we converted accented characters to their unaccented forms and transformed all text into lowercase letters. We implemented this preprocessing stage in Python, using the \emph{pandas, re, nltk, BeautifulSoup}, and \emph{unidecode} libraries.

\subsubsection{Stage V - Topic Modeling} \label{subsubsection_mod_top_I}

We developed this stage with two main objectives: identifying and marking posts containing cybersecurity-related keywords from our predefined list and identifying contextually irrelevant words to treat them as new stopwords. We conducted this analysis using LDA topic modeling, which helped us identify keywords and organize the data into topics. This organization enabled us to analyze the context of posts with and without keywords more deeply.

Similar to our IoC detection process described in Subsection \ref{subsubsection_extra_iocs}, we will use the presence of keywords as a parameter in our labeling process, which we detail in Subsection \ref{subsubsection_rotulagem}. Our system uses an attribute named \emph{KEYWORD} to track keyword presence in posts. It initially sets this attribute to \emph{NO} and updates it to \emph{YES} when it identifies at least one keyword from our predefined list.

Table \ref{tab_lista_de_keywords} shows our list of search keywords. We based some of these keywords on the work of \cite{deliu2018collecting}, while we added others based on our dataset's context.

\begin{table}[!ht]
\centering
\caption{List of keywords considered relevant in the context of cybersecurity}
\label{tab_lista_de_keywords}
\begin{tabular}{p{12.0cm}}
\hline
\multicolumn{1}{c}{\textbf{Considered Keywords}} \\ \hline
cpf, cpfs, cve, password, passwords, senha, senhas, hack, hacker, hackers, hacking, virus, malware, spyware, phishing, fishing, spam, trojan, criptografia, 
rootkit, backdoor, worm, botnet, vazamento, vazamentos, dados, spoofing, wordlist, ransomware, injection, sqlinjection, ddos, exploit, keylogger, vulnerabilidade, 
vulnerabilidades, hash, hashes \\ \hline
\end{tabular}
\end{table}

Table \ref{tab_lista_de_novas_stopwords} shows an example of words that, after analysis using LDA topic modeling, were considered new stopwords and, therefore, removed from the text. It is important to emphasize that LDA can be run multiple times with different predefined numbers of topics, aiming to identify the most significant possible number of new stopwords.

\begin{table}[!ht]
\centering
\caption{New stopwords found and removed from the text}
\label{tab_lista_de_novas_stopwords}
\begin{tabular}{p{12.0cm}} 
\hline
\multicolumn{1}{c}{\textbf{New stopwords found}} \\ \hline
pra, etc, none, vai, ter, nan, user, author, title, none, name, score, content, down, votes, created, comments, comment, answercontent, vote, type, points, aqui, pode, sobre, fazer, 
alguem, tudo, regular, coisa, bem, vou, sei, boca, algum, alguns, alguma, algo, nada, bom, entao, acho, quer, the, and, you, cara, coisas, sim, ainda, ver, usar, assim, index \\ \hline
\end{tabular}
\end{table}

\subsubsection{Stage VI - Labeling} \label{subsubsection_rotulagem}

Developing supervised machine learning models requires a labeled dataset. While researchers can sometimes use public datasets to train models and simplify the process, we found this approach unfeasible due to the nature of our data source -- the Dark Web -- which researchers still need to explore thoroughly to identify security incidents. Given this challenge and considering the study's specificity, we built our own labeled dataset. This decision represents a significant contribution to the information security community, as we will make the dataset available upon request.

We began the data labeling stage by simultaneously identifying posts containing IoCs and keywords. We categorized posts as \emph{Relevant} when they contained at least one IoC and one keyword. Conversely, we labeled posts \emph{Not Relevant} when both elements are missing. We flagged posts containing either an IoC or a keyword (but not both) for additional analysis. We named this initial labeled dataset \emph{DATASET I}.

Since data labeling requires meticulous evaluation, we used this initial dataset only to assess the preliminary performance of the machine learning algorithms. For the final labeling, we manually analyzed all posts and revised our initial labels, which relied solely on the presence of IoCs and keywords. Our analysis now considered the complete post content and additional characteristics, such as the post category. We expanded the final version of the labeled dataset to include all 26,575 initial posts. During this process, we reassessed and recategorized posts containing only IoCs or keywords we had previously removed, marking each as \emph{Relevant} or \emph{Not Relevant}. We named this final version \emph{DATASET II}.

\subsection{Development of the Post Classification Model and Testing}\label{subsection_desenvol_mod_classificacao}

With the labeled data, the work progressed to the development phase of the post classification model, followed by the testing phase to identify relevant or potentially malicious posts in new data collected from the Dark Web.

\begin{figure}[!ht]
    \center
    \includegraphics[width=1\textwidth]{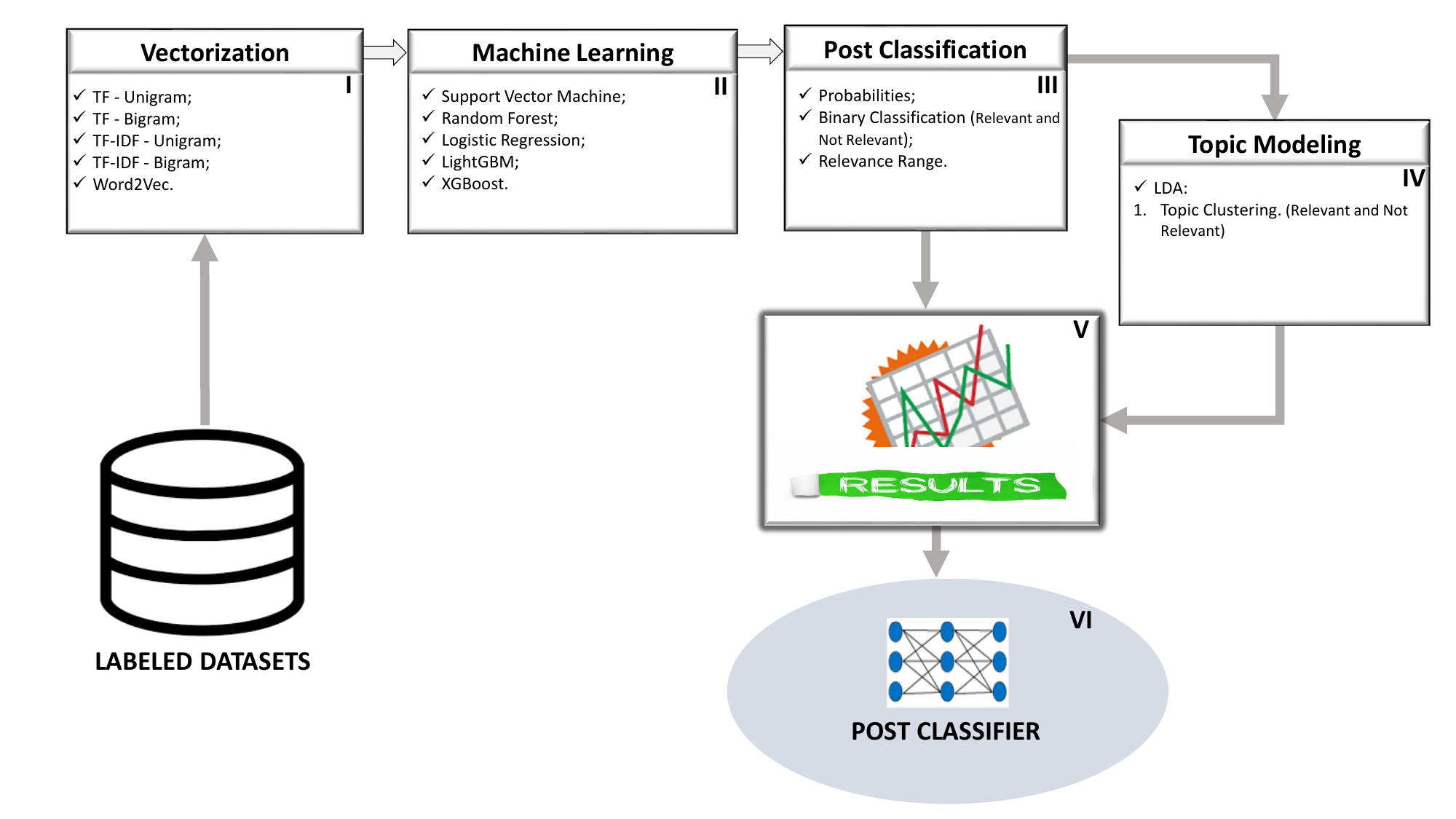}
    \caption{Stages of the post classification model development phase}
    \label{fig_proposta_modelo}
\end{figure}

Figure \ref{fig_proposta_modelo} illustrates our post classification model development stages. After careful preprocessing, the process begins with text vectorization, followed by applying machine learning algorithms, classification tasks, topic modeling, and finally, results analysis.

\subsubsection{Stage I - Vectorization} \label{subsubsection_vetorizacao}

The vectorization step transforms text into numerical vectors through two essential processes: feature extraction and token generation. We represented the text using five techniques: \emph{TF Unigram, TF Bigram, TF-IDF Unigram, TF-IDF Bigram,} and \emph{Word2Vec}.

These different representations ensure that the data is ready to be processed by machine learning algorithms, allowing for deeper analysis and information extraction. Some related works, such as \cite{dong2018new}, \cite{queiroz2019detecting}, \cite{samtani2020proactively}, and \cite{koloveas2021intime}, have cited the use of some of these representations, mainly involving Word2Vec and TF-IDF. However, during the investigation, no previous work was found to have tested all these representations.

The decision to explore various text representations was motivated by each vectorization technique's unique ability to capture specific information from the text. While TF and TF-IDF approaches emphasize the significance of term frequency and inverse term frequency, the Word2Vec method stands out by considering the semantic context. The underlying goal of this variety of approaches was to identify which fits the context most effectively.

\subsubsection{Stage II - Machine Learning} \label{subsubsection_aprendizado}

In this step, we evaluated five supervised machine learning algorithms to identify the best performer: \emph{Support Vector Machine}, \emph{Random Forest}, \emph{Logistic Regression}, \emph{LightGBM}, and \emph{XGBoost}. We maintained most algorithm parameters at their default values, making only specific adjustments when necessary.

We selected these algorithms based on their proven performance in related literature. Several researchers have successfully applied these methods: Deliu et al. \cite{deliu2018collecting}, Dong et al. \cite{dong2018new}, Queiroz et al. \cite{queiroz2019detecting}, and Koloveas et al. \cite{koloveas2021intime} showed SVM's effectiveness; Koloveas et al. \cite{koloveas2021intime} validated Random Forest's capabilities; and Sarkar et al. \cite{sarkar2019predicting} and Koloveas et al. \cite{koloveas2021intime} confirmed Logistic Regression's value. While the related studies did not specifically address LightGBM and XGBoost, we included them for their recognized advantages in efficiency, training time, and handling of imbalanced classes. Our literature review in Section \ref{sec_background} highlighted these benefits. Moreover, data science professionals frequently use these algorithms in Kaggle competitions, an online platform that hosts challenges, datasets, and resources for machine learning practitioners \cite{bojer2021kaggle}.

We first tested our models using an 80-20 train-test split ratio and later experimented with a 90-10 split. Evaluating both configurations helped us identify the most effective strategy for our application while enabling comparative analysis.

\subsubsection{Stage III - Post Classification} \label{subsubsection_classificacao}

We began our analysis with \emph{DATASET I}, described in Subsection \ref{subsubsection_rotulagem}, by testing the five algorithms from Subsection \ref{subsubsection_aprendizado} against the five data representations outlined in Subsection \ref{subsubsection_vetorizacao}. This process resulted in 25 unique algorithm-representation combinations. From these, we selected the models that achieved performance above 60\% across all key metrics (precision, recall, and F-score) for further training using \emph{DATASET II}, which comprises our complete labeled dataset containing all collected posts.

\begin{figure}[!ht]
    \center
    \includegraphics[width=1\textwidth]{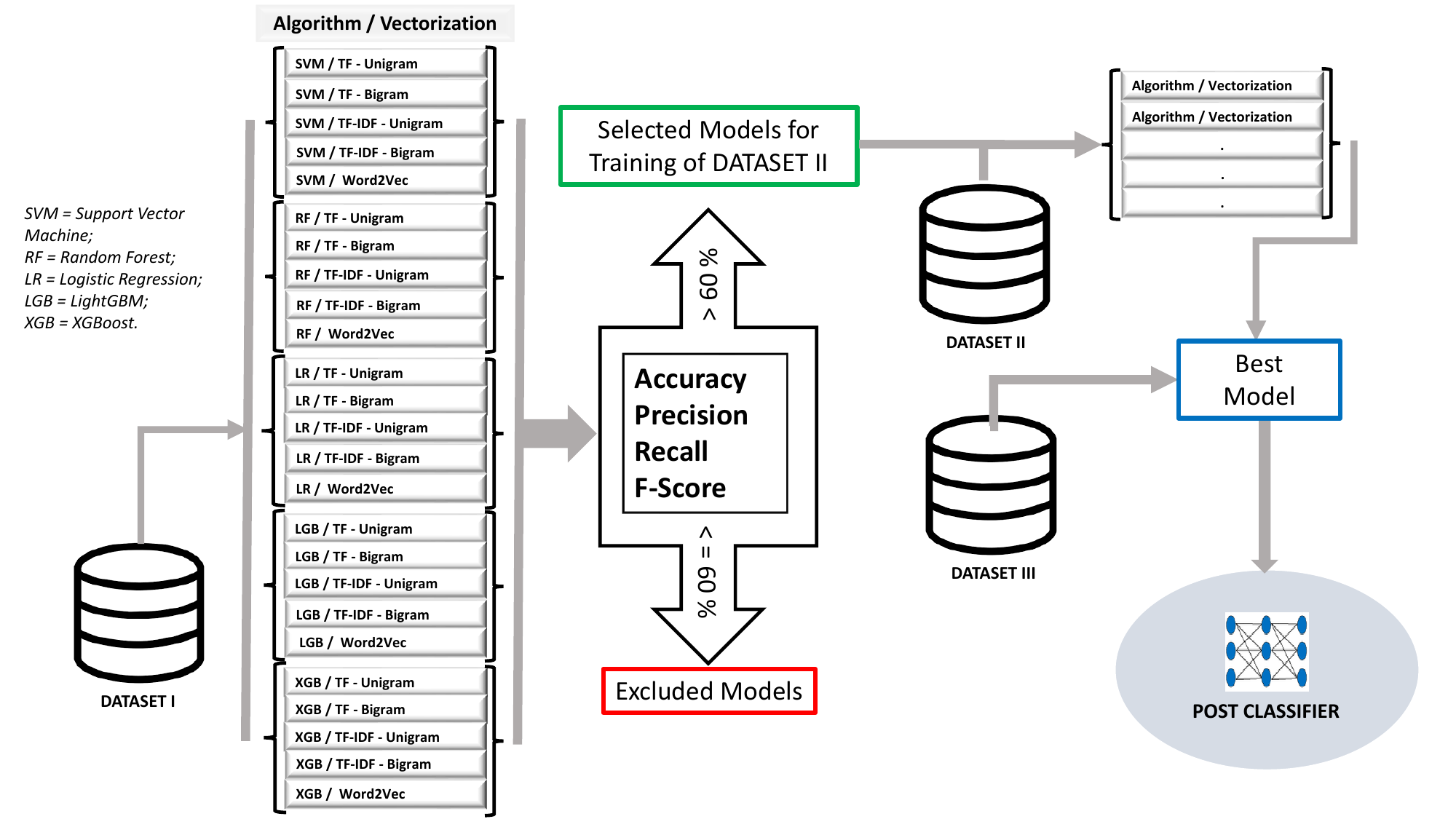}
    \caption{Post classification process flow}
    \label{fig_fluxo_processo_de_classificacao}
\end{figure}

We selected the best-performing model for post classification based on our training results. Our validation process involved two steps: First, we validated the model's performance on our existing labeled dataset. Second, to evaluate how well the model could classify new, unseen data, we created DATASET III: a new collection of 7,498 posts from the same forums not used in the training phase. Figure \ref{fig_fluxo_processo_de_classificacao} presents a complete flowchart of our classification process, including how we selected the optimal classification model.

We considered an output that indicated the probability of each post being relevant, with values between 0 and 1. A higher probability indicated greater confidence that the post was relevant. We implemented two classification approaches:

\begin{enumerate}
    \item \textbf{Binary Classification:}
    \begin{itemize}
        \item Probability $<$ 0.5: \emph{Not Relevant}
        \item Probability $\geq$ 0.5: \emph{Relevant}
    \end{itemize}

    \item \textbf{Three-Level Classification:}
    \begin{itemize}
        \item Probability $<$ 0.3: \emph{Low Relevance}
        \item $0.3 \leq$ Probability $\leq 0.7$: \emph{Medium Relevance}
        \item Probability $>$ 0.7: \emph{High Relevance}
    \end{itemize}
\end{enumerate}

\subsection{Stage IV - Topic Modeling} \label{subsection_mod_top_II}

After classifying the posts as \emph{Relevant} or \emph{Not Relevant}, we applied topic modeling using LDA on DATASETS II and III. Our topic modeling process followed these steps:

\begin{enumerate}
    \item Initial Clustering:
    \begin{itemize}
        \item Generated 20 topics from all posts
        \item Generated 10 topics from all posts
    \end{itemize}
    
    \item Separate Analysis by Relevance:
    \begin{itemize}
        \item Generated 10 topics from \emph{Relevant} posts only
        \item Generated 10 topics from \emph{Not Relevant} posts only
    \end{itemize}
\end{enumerate}

Each topic comprises documents, with words distributed according to their probability of occurrence. Table \ref{tab_mod_topicos} provides detailed information about the topic distribution across these different clustering approaches.

\begin{table}[!ht]
\centering
\caption{Topic clustering performed on datasets II and III using LDA}
\label{tab_mod_topicos}
\begin{tabular}{ccl}
\hline
\textbf{Dataset} & \textbf{Number of Topics} & \textbf{Coverage} \\ \hline
\multirow{4}{*}{DATASET II} & 20 & All posts \\ \hhline{~--}
                                      & 10 & All posts \\ \hhline{~--}
                                      & 10 & Posts Not Relevant \\ \hhline{~--}
                                      & 10 & Posts Relevant \\ \hline
\multirow{4}{*}{DATASET III} & 20 & All posts \\ \hhline{~--}
                                       & 10 & All posts \\ \hhline{~--}
                                       & 10 & Posts Not Relevant \\ \hhline{~--}
                                       & 10 & Posts Relevant \\ \hline
\end{tabular}
\end{table}

The topic clustering performed on both the labeled dataset \emph{DATASET II} and the unlabeled dataset \emph{DATASET III}, which was classified using the best-trained classification model, aimed to assess the similarity between the topics of each set and, consequently, to obtain an overview of the model's performance, even without the labels of the samples. For example, if the topics found in \emph{DATASET II} were similar to those found in \emph{DATASET III}, our confidence in the model's performance increases.

\subsubsection{Stage V - Results} \label{subsubsection_resultados}

This stage presents and discusses the results obtained from the previous stages. Regarding the supervised machine learning algorithms, the performance metrics were calculated, along with the analysis of the confusion matrices generated for each trained model.

We analyzed the most significant words from each topic to identify their key characteristics. By comparing topics generated from both labeled and unlabeled datasets, we gained more profound insights into how our model classifies previously unseen data.

\subsection{Identification of Relevant Posts in New Data Collected from the Dark Web} \label{subsubsection_testes_ident_novos_posts}

\begin{figure}[htb]
    \center
    \includegraphics[width=1\textwidth]{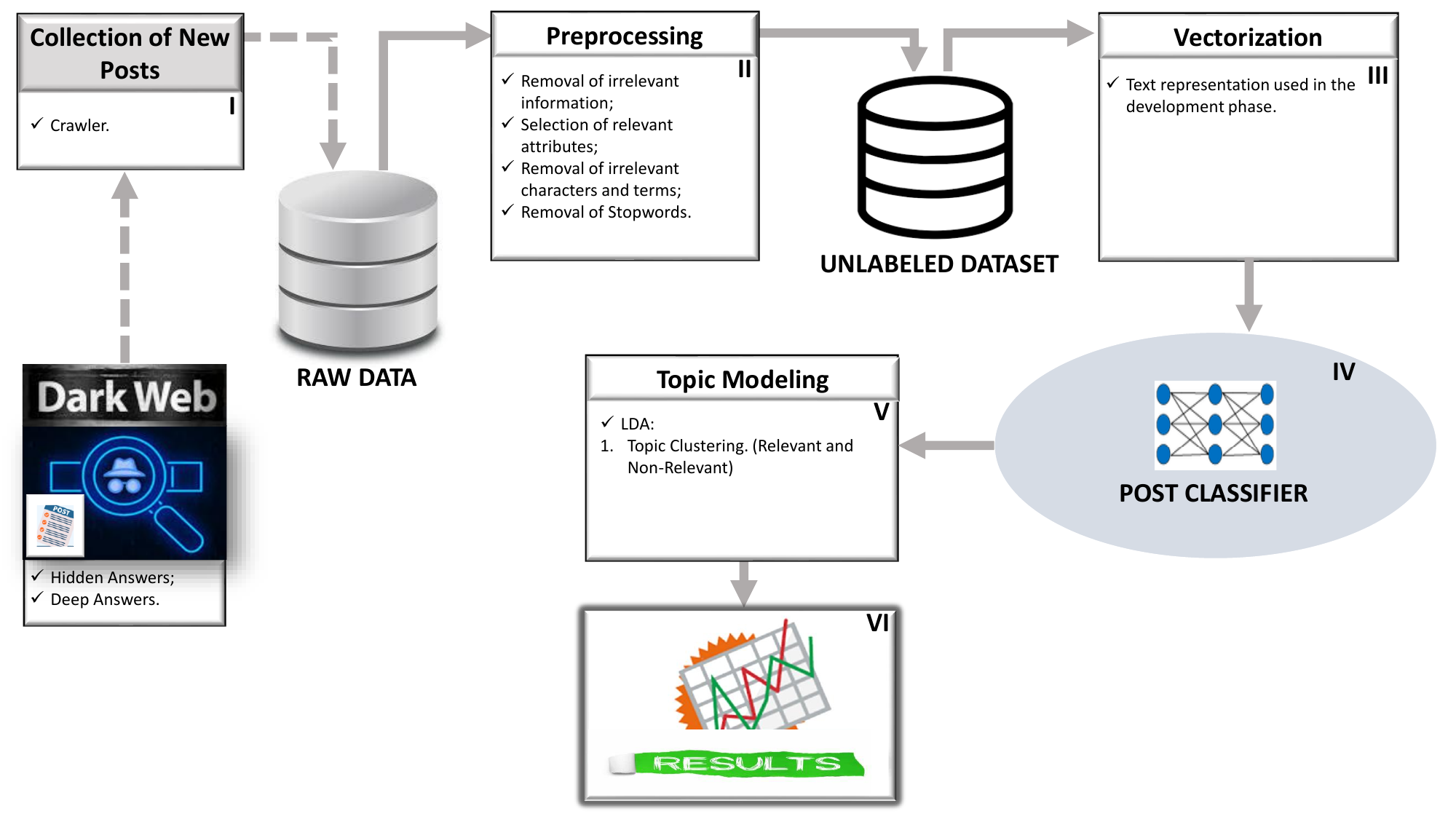}
    \caption{Steps of the testing phase for the model Identification of relevant posts in new data collected from the Dark Web}
    \label{fig_proposta_ident_novos_posts}
\end{figure}

We evaluated our post classifier using an unlabeled dataset of new posts not seen during training, as described in Subsection \ref{subsubsection_classificacao}. Figure \ref{fig_proposta_ident_novos_posts} illustrates the classifier testing process.

The process follows these stages. First, we collect and preprocess new posts to create \emph{DATASET III}. Then, we vectorize the text using the same pattern from the development phase and apply our classifier. Finally, we perform topic modeling on both classes and compare these topics with those from the labeled dataset to analyze the results.

\section{Experiments and Results}
\label{sec_resultados}

This section details the experiments following the method proposed Section \ref{sec_materiais_metodos}. Subsection \ref{subsection_conjuntos_de_dados} provides details about the datasets, Subsection \ref{subsection_selecao_classificadores} outlines the process for selecting the best classification model, Subsection \ref{subsec_melhores_classificadores} presents the results of the best classifier tests, Subsection \ref{subsec_testes_do_modelo_novos_posts} shows the model's performance in classifying new posts, Subsection \ref{subsec_resultados_lda} discusses the analysis of LDA topics, Subsection \ref{subsec_resultados_palavras_mais_freq} displays the analysis of the most frequent words in the \emph{Relevant} and \emph{Non-Relevant} post classes from the datasets. Finally, Subsection \ref{subsec_comp_trabalhos_rel} compares this article with related works.

\subsection{Datasets} \label{subsection_conjuntos_de_dados}

As mentioned in Section \ref{sec_materiais_metodos}, we analyzed data from two Dark Web foruns: \emph{Hidden Answers} and \emph{Deep Answer}. We collected using a crawler developed in the \emph{Go} programming language. For training the supervised machine learning models, we used 26,575 posts, which we labeled, resulting in \emph{DATASET I} and \emph{DATASET II}, detailed in Subsection \ref{subsubsection_processo_rotulagem}.

Table \ref{tab_coleta_i} presents the details of the dataset, including the number of posts, the posting period, and the language of the messages from the two forums that comprise the datasets. The \emph{Hidden Answers} forum shows a gap in the posting periods due to a period of forum inactivity.

\begin{table}[!ht]
\centering
\caption{Details of the posts collected for the training dataset of the supervised machine learning models}
\label{tab_coleta_i}
\begin{tabular}{lllcl}
\hline
\multicolumn{2}{l}{\textbf{Forum}}          & \textbf{Posting Period}                    & \multicolumn{1}{l}{\textit{\textbf{Posts}}} & \multicolumn{1}{l}{\textbf{Language}}     \\ \hline
\multicolumn{2}{l}{\textit{Hidden Answers}} & From 11/26/2016 to 04/12/2021                 & 19,652                                       & \multicolumn{1}{l}{Brazilian Portuguese} \\ \hline
\multicolumn{2}{l}{\textit{Hidden Answers}} & From 07/31/2021 to 07/15/2022                 & 6,681                                        & \multicolumn{1}{l}{Brazilian Portuguese} \\ \hline
\multicolumn{2}{l}{\textit{Deep Answers}}   & From 08/24/2021 e 09/14/2022                 & 242                                          & \multicolumn{1}{l}{Brazilian Portuguese} \\ \hline
                       &                      & \multicolumn{1}{r}{\textbf{Total posts:}} & \textbf{26,575}                              &                                          \\ \hhline{~~~|-|}
\end{tabular}
\end{table}

We performed the classification model with 7,498 new posts extracted from the same forums, as detailed in Table \ref{tab_coleta_ii}. We presented these previously unseen messages to the model to evaluate its performance. As mentioned in Subsection \ref{subsubsection_classificacao}, we designated this set of new unlabeled posts as \emph{DATASET III}.

\begin{table}[!ht]
\centering
\caption{Details of the posts collected for the test database of the post classification model}
\label{tab_coleta_ii}
\begin{tabular}{lllcl}
\hline
\multicolumn{2}{l}{\textbf{Forum}}          & \textbf{Posting Period}                    & \multicolumn{1}{l}{Posts} & \multicolumn{1}{l}{\textbf{Language}}     \\ \hline
\multicolumn{2}{l}{\emph{Hidden Answers}}   & From 09/10/2022 to 07/10/2023                 & 7,343                                       & \multicolumn{1}{l}{Brazilian Portuguese} \\ \hline
\multicolumn{2}{l}{\emph{Deep Answers}}     & From 09/16/2022 to 01/01/2023                 & 155                                          & \multicolumn{1}{l}{Brazilian Portuguese} \\ \hline
                       &                      & \multicolumn{1}{r}{\textbf{Total posts:}} & \textbf{7,498}                              &                                          \\ \hhline{~~~|-|}
\end{tabular}
\end{table}

\subsubsection{Dataset labeling process} \label{subsubsection_processo_rotulagem}

We conducted the post-labeling process using two different approaches as described in Subsection \ref{subsubsection_rotulagem}. We considered the simultaneous occurrence of \emph{IoCs} and \emph{keywords} in the first approach. In the second approach, we analyzed the posts manually, considering not only the occurrence of \emph{IoCs} and \emph{keywords} but also the content and other characteristics, such as the category.

In the first approach, among the 26,575 posts detailed in Table \ref{tab_coleta_i}, 16,010 did not contain any IoC or keywords, so we marked them as \emph{Not Relevant}. We found IoCs in 6,926 posts and keywords in 5,304 posts. The intersection between posts containing both \emph{IoCs} and \emph{keywords} totaled 1,665 posts, which we marked as \emph{Relevant}.

This approach resulted in \emph{DATASET I}, our first labeled database for training supervised machine learning models. This database contains 17,675 posts, with 1,665 (approximately 9\%) labeled as \emph{Relevant} and 16,010 (about 91\%) as \emph{Not Relevant}.

In the second approach, we analyzed all 26,575 posts, including those containing only IoCs or only keywords that we had excluded in the first labeling stage. We evaluated these posts and labeled them as either \emph{Relevant} or \emph{Not Relevant}. We named this final version \emph{DATASET II}, which contains 3,341 posts (approximately 13\%) labeled as \emph{Relevant} and 23,234 posts (about 87\%) as \emph{Not Relevant}.

Table \ref{tab_conjuntos_de_dados} summarizes the datasets used in this research. \emph{DATASETS I} and \emph{II} contain the same posts detailed in Table \ref{tab_coleta_i}, differing only in their labeling approach. \emph{DATASET III} consists of entirely new posts we collected specifically for testing the classification model, as mentioned earlier in this section. The labeled datasets (\emph{DATASET I} and \emph{DATASET II}) present an imbalanced distribution, with fewer posts classified as relevant compared to non-relevant posts. To address this, we prioritized evaluation metrics that are robust to class imbalance, particularly the F1-score. No explicit sampling techniques (e.g., SMOTE or undersampling) were applied to preserve the natural distribution of real-world data.

\begin{table}[!ht]
\centering
\caption{Details of the datasets used in the development of the research}
\label{tab_conjuntos_de_dados}
\begin{tabular}{lllll}
\hline
\multicolumn{1}{l}{\textbf{Dataset}}  & \multicolumn{1}{l}{\textbf{Total Posts}} & \multicolumn{3}{c}{\textbf{Label}}                                                            \\ \hline
\multicolumn{1}{l}{\multirow{2}{*}{DATASET I}}  & \multicolumn{1}{c}{\multirow{2}{*}{17,675}} & \multicolumn{1}{l}{Relevant}     & \multicolumn{1}{l}{1,665}  & \multicolumn{1}{l}{9.42\%}  \\ \hhline{~~~|-|-|}
\multicolumn{1}{l}{}                            & \multicolumn{1}{c}{}                        & \multicolumn{1}{l}{Not Relevant} & \multicolumn{1}{l}{16,010} & \multicolumn{1}{l}{90.58\%} \\ \hline
                                                  &                                              &                                    &                             &                              \\ \hline
\multicolumn{1}{l}{\textbf{Dataset}}  & \multicolumn{1}{l}{\textbf{Total Posts}} & \multicolumn{3}{c}{\textbf{Label}}                                                            \\ \hline
\multicolumn{1}{l}{\multirow{2}{*}{DATASET II}} & \multicolumn{1}{c}{\multirow{2}{*}{26,575}} & \multicolumn{1}{l}{Relevant}     & \multicolumn{1}{l}{3,341}  & \multicolumn{1}{l}{12.57\%} \\ \hhline{~~~|-|-|}
\multicolumn{1}{l}{}                            & \multicolumn{1}{c}{}                        & \multicolumn{1}{l}{Not Relevant} & \multicolumn{1}{l}{23,234} & \multicolumn{1}{l}{87.43\%} \\ \hline
                                                  &                                              &                                    &                             &                              \\ \hline
\multicolumn{1}{l}{\textbf{Dataset}}  & \multicolumn{1}{l}{\textbf{Total Posts}} & \multicolumn{3}{c}{\textbf{Label}}                                                            \\ \hline
\multicolumn{1}{l}{DATASET III}                 & \multicolumn{1}{c}{7,498}                   & \multicolumn{3}{c}{Without Label}                                                                 \\ \hline
\end{tabular}
\end{table}

\subsection{Selection of the Best Classification Model} \label{subsection_selecao_classificadores}

To identify the best classifier, we evaluated five classification algorithms with five different data representations. We split the data into 80\% for training and 20\% for testing. We used \emph{DATASET I} for this testing phase.

\begin{figure}[!ht]
    \center
    \includegraphics[width=1\textwidth]{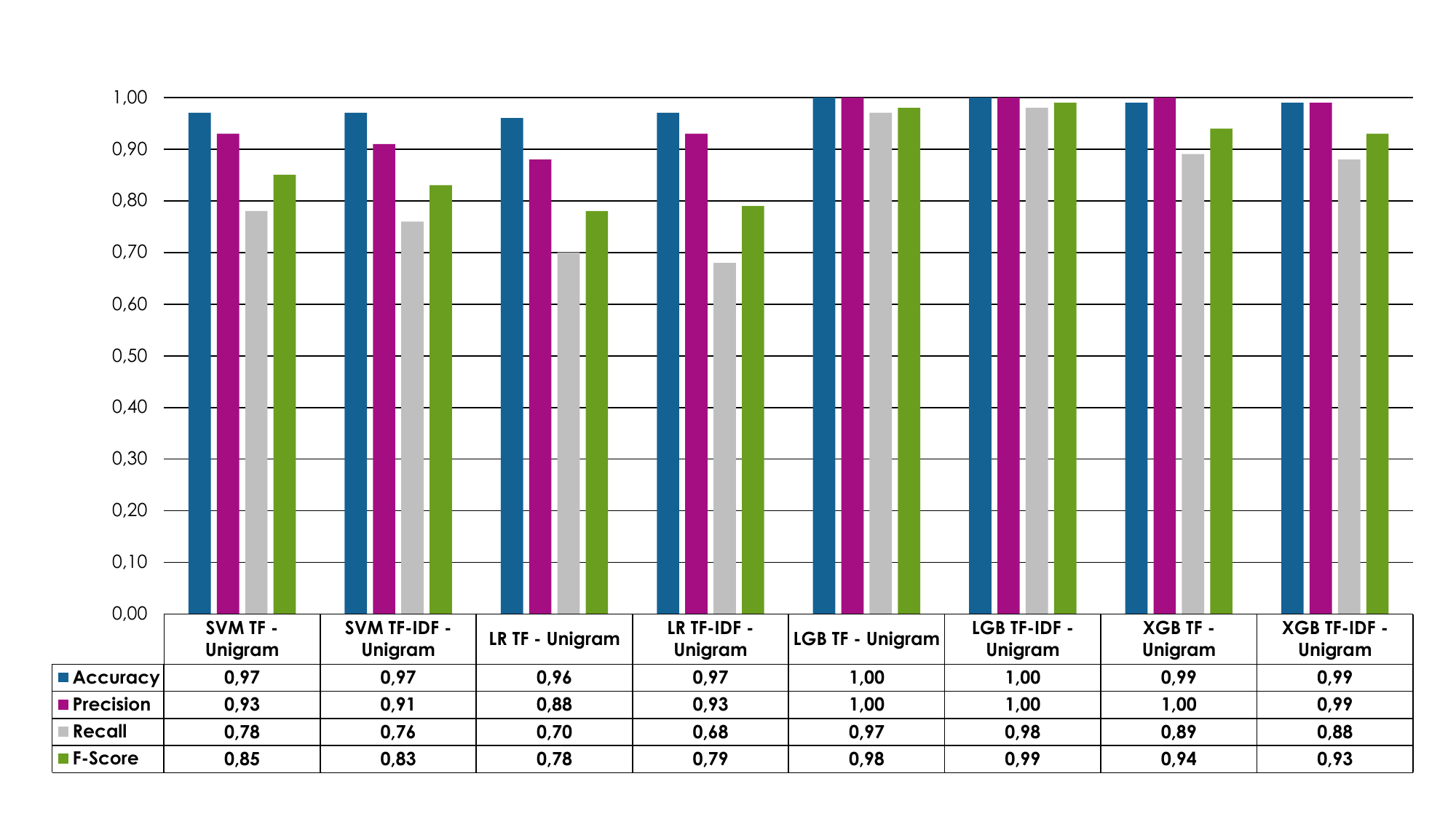}
    \caption {Performance metrics of the best classifiers evaluated on the \emph{DATASET I}}
    \label{fig_metricas_desempenho_melhores_classificadores}
\end{figure}

From the 25 tested combinations (five learning algorithms: SVM, Random Forest, Logistic Regression, LightGBM, and XGBoost and five text representation methods: TF - Unigram, TF - Bigram, TF-IDF - Unigram, TF-IDF - Bigram, and Word2Vec), we selected those with precision, recall, and F-score metrics above 60\%. The combinations of SVM, Logistic Regression, LightGBM, and XGBoost algorithms with TF-Unigram and TF-IDF-Unigram representations exceeded our minimum metric thresholds. Table \ref{tab_melhores_modelos_classificacao} shows these top-performing combinations, while Figure \ref{fig_metricas_desempenho_melhores_classificadores} displays the performance metrics for each classifier for class 1 (Relevant Posts).

\begin{table}[!ht]
\centering
\caption{Supervised machine learning algorithms and text representations that achieved metrics above 60\%}
\label{tab_melhores_modelos_classificacao}
\begin{tabular}{lccccc}
\hline
\multicolumn{1}{c}{\multirow{2}{*}{\textbf{\begin{tabular}[c]{@{}c@{}}Text Vector \\ Representation \end{tabular}}}} & \multicolumn{5}{c}{\textbf{Supervised Machine Learning Algorithms}}                                                                                                                                                                                                                     \\ \hhline{~-----} 
\multicolumn{1}{c}{}                                                                                                     & \multicolumn{1}{c}{\begin{tabular}[c]{@{}c@{}}Support Vector\\  Machine\end{tabular}} & \multicolumn{1}{c}{\begin{tabular}[c]{@{}c@{}}Random\\ Forest\end{tabular}} & \multicolumn{1}{c}{\begin{tabular}[c]{@{}c@{}}Logistic\\  Regression\end{tabular}} & \multicolumn{1}{c}{LightGBM} & XGBoost \\ \hline
TF - Unigram                                                                                                               & \multicolumn{1}{c}{\textcolor{green}{\checkmark}}                                                                 & \multicolumn{1}{c}{}                                                        & \multicolumn{1}{c}{\textcolor{green}{\checkmark}}                                                              & \multicolumn{1}{c}{\textcolor{green}{\checkmark}}        & \textcolor{green}{\checkmark}       \\ \hline
TF - Bigram                                                                                                                & \multicolumn{1}{c}{}                                                                  & \multicolumn{1}{c}{}                                                        & \multicolumn{1}{c}{}                                                               & \multicolumn{1}{c}{}         &         \\ \hline
TF-IDF - Unigram                                                                                                           & \multicolumn{1}{c}{\textcolor{green}{\checkmark}}                                                                 & \multicolumn{1}{c}{}                                                        & \multicolumn{1}{c}{\textcolor{green}{\checkmark}}                                                              & \multicolumn{1}{c}{\textcolor{green}{\checkmark}}        & \textcolor{green}{\checkmark}       \\ \hline
TF-IDF - Bigram                                                                                                            & \multicolumn{1}{c}{}                                                                  & \multicolumn{1}{c}{}                                                        & \multicolumn{1}{c}{}                                                               & \multicolumn{1}{c}{}         &         \\ \hline
Word2Vec                                                                                                                   & \multicolumn{1}{c}{}                                                                  & \multicolumn{1}{c}{}                                                        & \multicolumn{1}{c}{}                                                               & \multicolumn{1}{c}{}         &         \\ \hline
\end{tabular}
\end{table}

The results are highly satisfactory, with most metrics exceeding 90\% and some even reaching 100\%. Only a few values fall within the 70\% range. These results are consistent with expectations, as the labeling process adopted for \emph{DATASET I} is believed to have effectively segmented the posts into two distinct classes. Overall, the LightGBM algorithm using TF-IDF - Unigram achieved the best performance.

\subsection{Testing the Best Classification Model} \label{subsec_melhores_classificadores}

The best classification models, selected as described in Subsection \ref{subsection_selecao_classificadores}, were trained using \emph{DATASET II}. The goal was to identify the best post classification model among the eight combinations (learning algorithm/text representation) that presented the highest performance metrics in the previous experiment.

\begin{figure}[!ht]
    \center
    \includegraphics[width=1\textwidth]{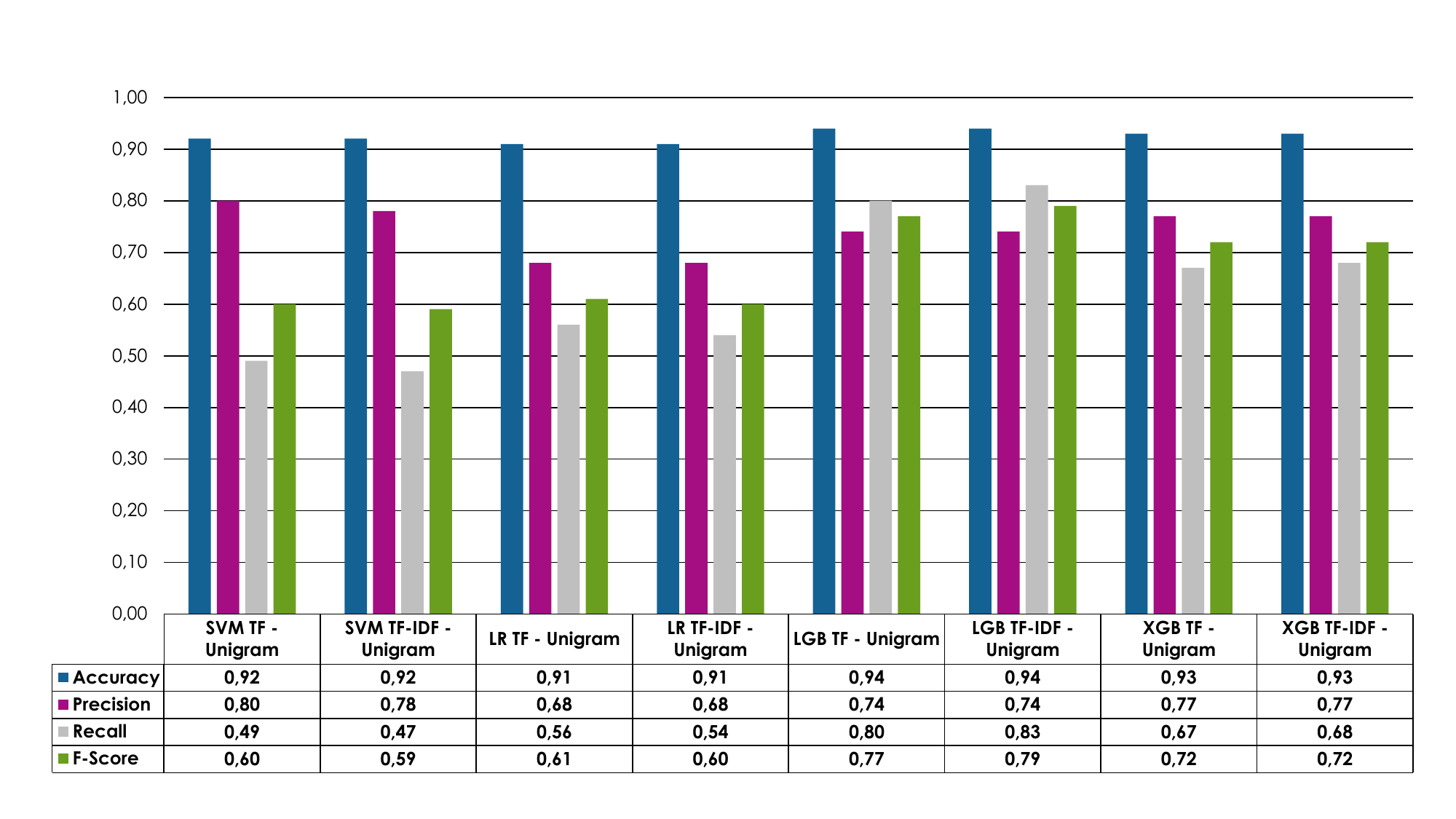}
    \caption{Performance metrics of the best classifiers evaluated on \emph{DATASET II}}
    \label{fig_metricas_desempenho_melhores_classificadores_DT2}
\end{figure}

In the first stage of this training, we split the data, allocating 80\% for training and 20\% for testing. Subsequently, we tested the best-performing model with a 90\% training and 10\% testing split. This change did not result in significant differences in the results. Figure \ref{fig_metricas_desempenho_melhores_classificadores_DT2} shows the summary of results for class 1 \emph{(Relevant Posts)}, using the 80\% training and 20\% testing split. The results did not reach the same level achieved with the first dataset, which is an expected outcome given the adopted data labeling approach. Nevertheless, the values remain satisfactory.

\begin{figure}[!ht]
    \center
    \includegraphics[width=0.7\textwidth]{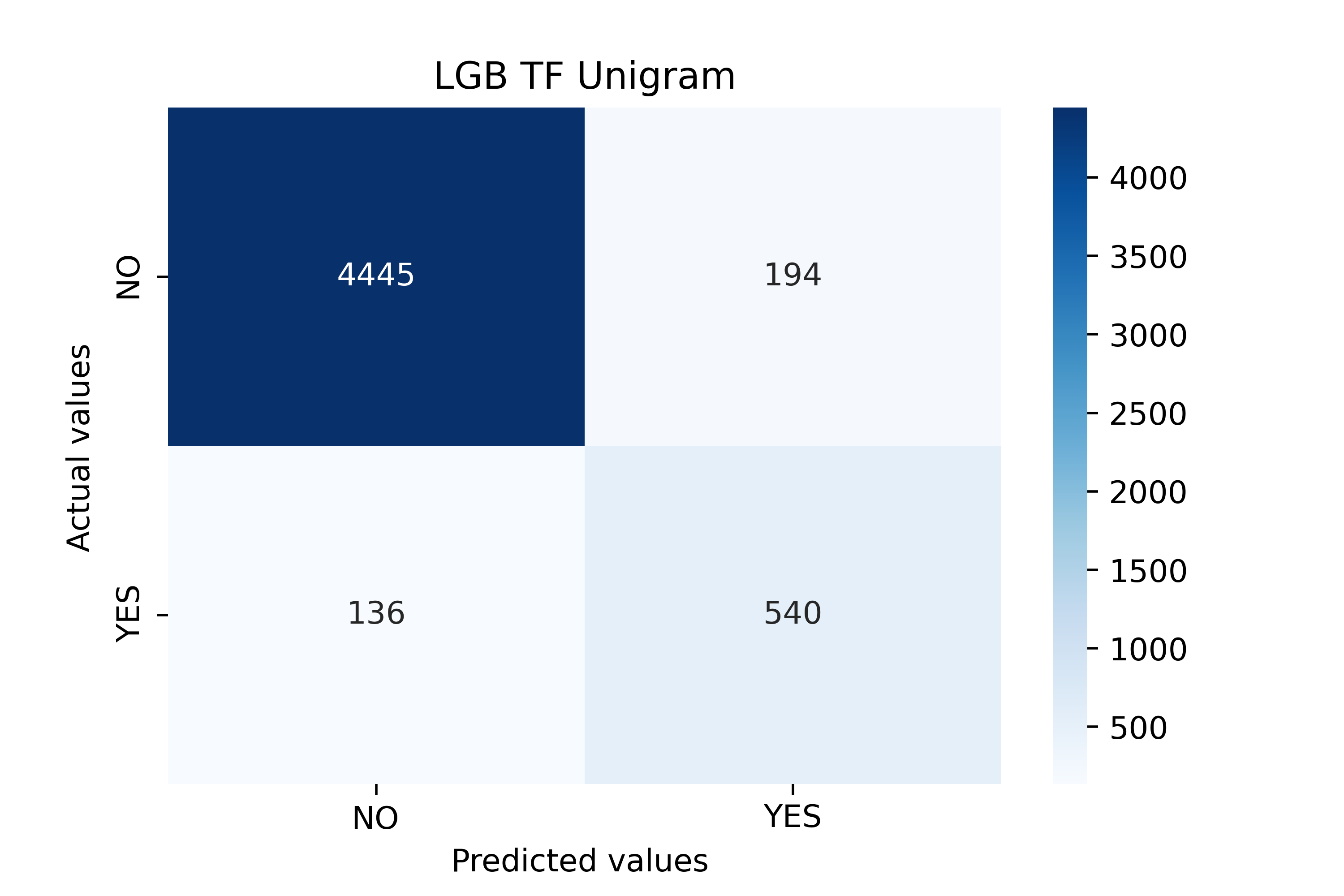}
    \caption{Confusion matrix of the LightGBM algorithm using TF - Unigram}
    \label{fig_matriz_lgb_tf_unigram}
\end{figure}

\begin{figure}[!ht]
    \center
    \includegraphics[width=0.7\textwidth]{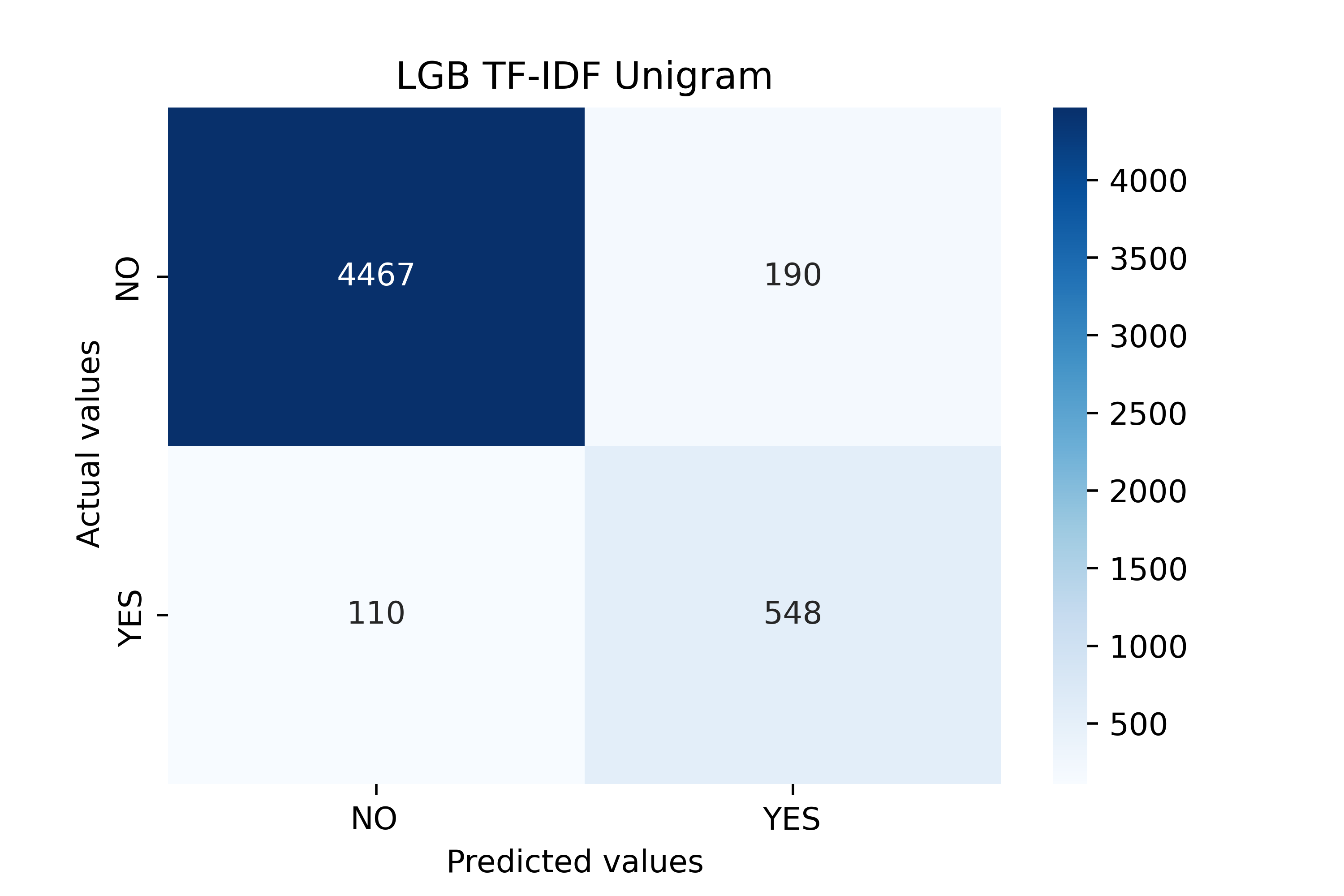}
    \caption{Confusion matrix of the LightGBM algorithm using TF-IDF - Unigram}
    \label{fig_matriz_lgb_tfidf_unigram}
\end{figure}

The \emph{LightGBM} algorithm, using the \emph{TF-IDF - Unigram} approach, once again delivered the best results, establishing itself as the best classification model to be adopted. Figures \ref{fig_matriz_lgb_tf_unigram} and \ref{fig_matriz_lgb_tfidf_unigram} illustrate the confusion matrices corresponding to each approach.

The results showed that the LightGBM algorithm performed well, which was characterized by a low number of false positives and a high accuracy rate. This performance was particularly notable when using the TF-IDF - Unigram approach, achieving an accuracy rate of 94\% in the \emph{Relevant} posts class.

\subsection{Testing the Model for Identifying Relevant Posts in New Unlabeled Data} \label{subsec_testes_do_modelo_novos_posts}

We tested the model for identifying relevant posts, implemented through the LightGBM algorithm and using the TF-IDF - Unigram text representation, on new unlabeled data, the \emph{DATASET III}. As described in Subsection \ref{subsubsection_classificacao}, the algorithm assigns a probability between 0 and 1 regarding the relevance of each analyzed post. During the testing phase, we applied two forms of classification: the first classifying posts as \emph{Relevant} or \emph{Not Relevant}, and the second considering relevance ranges \emph{(Low, Medium, or High)}.

The model classified the 7,498 posts in the dataset as follows: 1,158 (approximately 15\%) as \emph{Relevant} (potentially malicious messages) and 6,340 (approximately 85\%) as \emph{Not Relevant}.

A comparative analysis between the previously labeled \emph{DATASET II} and the classification results produced by the model on the \emph{DATASET III} reveals a proximity in values. In the \emph{DATASET II}, 13\% of the posts are marked as \emph{Relevant}, while the classification model identified 15\% of the posts in the \emph{DATASET III} as \emph{Relevant}.

The classification of 7,498 posts by relevance range showed 595 posts (8\%) as \emph{High Relevance}, 849 posts (11\%) as \emph{Medium Relevance}, and 6,054 posts (81\%) as \emph{Low Relevance}. This result indicates that a significant majority of the posts (81\%) do not contain relevant content concerning information security, a pattern also observed in the labeled dataset.

\subsection{Analysis of LDA Topic Modeling} \label{subsec_resultados_lda}

As described in Subsection \ref{subsection_mod_top_II}, an overview of the performance of the \emph{Relevant Posts} identification model was obtained by generating topics using the LDA technique on both datasets: the \emph{DATASET II} (labeled dataset) and the \emph{DATASET III}. Tables \ref{tab_20_topicos_todos_posts} and \ref{tab_20_topicos_todos_posts_novos_posts} present two of the 20 topics generated from all the posts in these datasets.

\begin{table}[!ht]
\centering
\caption{Two of the 20 Topics of all posts from the \emph{DATASET II}. The topic content was manually translated to english.}
\label{tab_20_topicos_todos_posts}
\begin{tabular}{cp{11.0cm}}
\hline
\multicolumn{1}{l}{\textbf{Topic}} & \multicolumn{1}{c}{\textbf{Main Words}}                                            \\ \hline
3      (pt-br)                               & brasil, pais, governo, guerra, estado, povo, paises, mundo, eua, poder                       \\ 

  3   (en)                                & brazil, country, government, war, state, people, countries, world, usa, power                       \\ \hline
8       (pt-br)                              & dados, nome, conta, pessoa, email, informacoes, telegram, conseguir, saber, cpf              \\ 
8         (en)                            & data, name, account, people, email, information, telegram, get, know, cpf  (social security number)            \\ \hline
\end{tabular}
\end{table}


\begin{table}[!ht]
\centering
\caption{Two of the 20 Topics of all posts from the \emph{DATASET III}. The topic content was manually translated to english.}
\label{tab_20_topicos_todos_posts_novos_posts}
\begin{tabular}{cp{11.0cm}}
\hline
\textbf{Topic} & \multicolumn{1}{c}{\textbf{Main Words}}                                       \\ \hline
2    (pt-br)           & brasil, governo, pais, povo, porque, bolsonaro, brasileiro, lula, desde, presidente     \\ 
2    (en)           & brazil, government, country, people, why, bolsonaro, brazilian, lula, since, president     \\ \hline
19   (pt-br)           & site, telegram, link, sites, dados, conta, comprar, links, onde, saber                  \\ 
19      (en)        & site, telegram, link, sites, data, account, buy, links, where, know                  \\ \hline
\end{tabular}
\end{table}


The 20 topics from the datasets provide a comprehensive view of their content. The main words of each topic reveal a wide variety of subjects discussed in the Dark Web forums. The datasets share some similarities. For example, topic 3 in Table \ref{tab_20_topicos_todos_posts} and topic 2 in Table \ref{tab_20_topicos_todos_posts_novos_posts} both contain words related to politics. Similarly, topic 8 in Table \ref{tab_20_topicos_todos_posts} and topic 19 in Table \ref{tab_20_topicos_todos_posts_novos_posts} share words related to data breaches.

We performed a second grouping to reduce the initial 20 topics to 10 for each dataset. This analysis revealed that we could still effectively identify the main subjects discussed even with fewer topics. The analysis also showed apparent similarities between the datasets, which was expected since they come from the same forums but cover different periods (as described in Section \ref{subsection_conjuntos_de_dados}).

Up to this point, topic modeling using LDA has revealed general topics discussed in the forums and highlighted similarities between the analyzed datasets. Next, to assess the model's ability to identify \emph{Relevant Posts}, topics were generated in both datasets, dividing them according to their labels. In this analysis, posts with a probability lower than 0.5 are classified as Not Relevant, while those with a probability equal to or greater than 0.5 are considered Relevant.

Tables \ref{tab_10_topicos_posts_nao_relevantes} and \ref{tab_10_topicos_posts_nao_relevantes_posts_novos} present some of the 10 topics generated in the datasets, considering only the \emph{Not Relevant} posts. On the other hand, Tables \ref{tab_10_topicos_posts_relevantes} and \ref{tab_10_topicos_posts_relevantes_posts_novos} display some the 10 topics generated by considering only the \emph{Relevant} posts.

\begin{table}[!ht]
\centering
\caption{Two of the 10 Topics of the \emph{Not Relevant} posts from the \emph{DATASET II}. The topic content was manually translated to english.}
\label{tab_10_topicos_posts_nao_relevantes}
\begin{tabular}{cp{11.0cm}}
\hline
\textbf{Topic} & \multicolumn{1}{c}{\textbf{Main Words}}                                    \\ \hline
6     (pt-br)          & dinheiro, comprar, ganhar, grana, vender, boa, facil, conta, compra, mercado         \\ 
6     (en)          & money, buy, win, money, sell, good, easy, account, buy, market         \\ \hline
8      (pt-br)         & brasil, pais, governo, guerra, estado, povo, paises, poder, contra, eua              \\ 
8       (en)        & brazil, country, government, war, state, people, countries, power, against, usa              \\ \hline
\end{tabular}
\end{table}


\begin{table}[!ht]
\centering
\caption{Two of the 10 Topics of the \emph{Not Relevant} posts from the \emph{DATASET III}. The topic content was manually translated to english.}
\label{tab_10_topicos_posts_nao_relevantes_posts_novos}
\begin{tabular}{cp{11.0cm}}
\hline
\textbf{Topic} & \multicolumn{1}{c}{\textbf{Main Words}}                              \\ \hline
3     (pt-br)          & dinheiro, brasil, mundo, lula, hoje, governo, sempre, sendo, fez, errado       \\ 
3     (en)          & money, brazil, world, lula, today, government, always, being, did, wrong       \\ \hline
10  (pt-br)            & dinheiro, melhor, faz, comprar, site, sabe, dar, saber, conta, boa             \\ 
10   (en)           & money, better, do, buy, site, know, give, know, account, good             \\ \hline
\end{tabular}
\end{table}


\begin{table}[!ht]
\centering
\caption{One of the 10 Topics of the \emph{Relevant} posts from the \emph{DATASET II}. The topic content was manually translated to english.}
\label{tab_10_topicos_posts_relevantes}
\begin{tabular}{cp{11.0cm}}
\hline
\textbf{Topic} & \multicolumn{1}{c}{\textbf{Main Words}}                                          \\ \hline
7       (pt-br)        & pessoa, numero, cpf, nome, dados, telegram, celular, social, engenharia, telefone          \\ 
7      (en)         & people, number, cpf (social security), name, data, telegram, cell phone, social, engineering, phone          \\ \hline
\end{tabular}
\end{table}


\begin{table}[!ht]
\centering
\caption{One of the 10 Topics of the \emph{Relevant} posts from the \emph{DATASET III}. The topic content was manually translated to english.}
\label{tab_10_topicos_posts_relevantes_posts_novos}
\begin{tabular}{cp{11.0cm}}
\hline
\textbf{Topic} & \multicolumn{1}{c}{\textbf{Main Words}}                                  \\ \hline
9    (pt-br)           & dados, cpf, numero, pessoa, telegram, cartao, site, nome, telefone, puxar          \\ 
9     (en)          & data, cpf (social security), number, people, telegram, card, site, name, phone, get          \\ \hline
\end{tabular}
\end{table}

Analyzing the topics generated from \emph{Non-Relevant} posts reveals similarities across the datasets. For example, ``politics'' appears in topic 8 of Table \ref{tab_10_topicos_posts_nao_relevantes} and in topic 3 of Table \ref{tab_10_topicos_posts_nao_relevantes_posts_novos}. Similarly, topics related to ``purchases'' are found in topic 6 of Table \ref{tab_10_topicos_posts_nao_relevantes} and topic 10 of Table \ref{tab_10_topicos_posts_nao_relevantes_posts_novos}. As expected, none of the topics address cyber threats, which are more likely to appear in the \emph{Relevant} posts class.

In contrast, almost all topics generated from \emph{Relevant} posts contain words directly related to cyber threats, with several indicating potential data leakage. For example, topic 7 of Table \ref{tab_10_topicos_posts_relevantes} and topic 9 of Table \ref{tab_10_topicos_posts_relevantes_posts_novos} highlight these themes. The similarities across topics in both datasets suggest that the model successfully learned to classify posts during the training phase, demonstrating its ability to identify and categorize new posts accurately.

Table \ref{tab_posts_real_prob_alta_cj_dados_iii} presents four posts classified as highly relevant in \emph{DATASET III}. Posts with IDs 899 and 1048 discuss data leaks involving individuals and companies, while posts with IDs 1010 and 6632 address software vulnerabilities. These findings illustrate the model's effectiveness in identifying posts with potentially valuable CTI information.

\begin{table}[!htp]
\caption{Example of posts identified as highly relevant in the \emph{DATASET III}}
\label{tab_posts_real_prob_alta_cj_dados_iii}
\centering
\begin{tabular}{lp{7cm}ll}
\hline
\textbf{ID} & \textbf{full\_text}                                                                                                                                                                                                                                                                                                                                                                                                                                                     & \textbf{created\_at} & \textbf{probability} \\ \hline
899         & ``record sabem site deep web vazaram dados record pastor enganava fieis obter dinheiro vazamento conhecimento ...  vazado sendo vendidos soubesse endereco onion ... hackers ... comecam divulgacao dados sensiveis roubados durante ataque vazamento parece maior avaliacao inicial documentos fotocopia passaporte ... planilhas detalhadas despesas receitas alem correspondencias internas departamento juridico empresa pastor enganava fieis obter dinheiro ''... & 10/17/2022           & 0,77194924             \\ \hline
1010        & ``vender vulnerabilidade tres meses achei vulnerabilidade risco medio tiktok ganhei recentemente writeup apple disse vender vulnerabilidades deep web ... falha empresa ... garantir confidencialidade pontas exploracao vulnerabilidade quanto compromete confidencialidade integridade disponibilidade diretamente ... vulnerabilidade poderia invadir contas tiktok poderia invadir servidor database tipo claramente ganharia agora apenas derrubar tiktok'' ...     & 10/23/2022           & 0,94460218             \\ \hline
1048        & ``indica plataforma boa vender dados forma segura possuo dados pessoais privados pessoas empresas vender gostaria encontrar plataforma segura vender mesmos atraves network telegram site google ... interesse dados completos endereco telefone cpf ... sabe acredito pagam pro telegram ... vender dados dados tipo cpf nome data nascimento onde mora tals site'' ...                                                                                                  & 10/24/2022           & 0,861122741            \\ \hline
6632        & ``vulnerabilidade isc bind explorando servidor achei isc bind versao queria tentar explorar vulnerabilidade nele achei exploits exploit queria saber trabalhou servidor trabalhou falhas qualquer ajuda dica vindanone''                                                                                                                                                                                                                                                  & 05/24/2023           & 0,820266759            \\ \hline
\end{tabular}
\end{table}


\subsection{Analysis of the Most Frequent Words in Each Class} \label{subsec_resultados_palavras_mais_freq}

Topic modeling revealed the dominant word groups in each generated topic. We analyzed the absolute frequency of the most common words to gain deeper insights. This quantitative analysis helped verify the similarities between the labeled data and the model's classifications. We enumerated the 100 most frequent words for each label, with the top 50 presented in Tables \ref{tab_lista_100_palavras_mais_freq_nao_rel} and \ref{tab_lista_100_palavras_mais_freq_rel}.


\begin{table}[!ht]
\caption{List of the 50 most frequent words in the class of not relevant posts from DATASETS II and III}
\label{tab_lista_100_palavras_mais_freq_nao_rel}
\centering
\begin{tabular}{p{6.2cm}|p{6.2cm}}
\hline
\multicolumn{2}{c}{\textbf{Most Frequent Words}} \\
\hline
\multicolumn{2}{c}{\textbf{Not Relevant Posts}} \\
\hline
\textbf{DATASET II} & \textbf{DATASET III} \\ \hline
1 - pessoas, 2 - vida, 3 - tempo, 4 - sempre, 5 - porque, 6 - dia, 7 - tipo, 8 - pessoa, 9 - melhor, 10 - nunca, 11 - faz, 12 - site, 13 - mundo, 14 - boa, 15 - apenas, 16 - mano, 17 - todos, 18 - pois, 19 - anos, 20 - todo, 21 - ficar, 22 - agora, 23 - qualquer, 24 - dinheiro, 25 - saber, 26 - forum, 27 - mim, 28 - forma, 29 - onde, 30 - sabe, 31 - dar, 32 - gente, 33 - outros, 34 - realmente, 35 - caso, 36 - menos, 37 - uns, 38 - vezes, 39 - hoje, 40 - vez, 41 - existe, 42 - quero, 43 - conta, 44 - pergunta, 45 - outro, 46 - pouco, 47 - falar, 48 - casa, 49 - merda, 50 - outra
& 1 - pessoas, 2 - vida, 3 - faz, 4 - tempo, 5 - porque, 6 - sempre, 7 - melhor, 8 - tipo, 9 - pessoa, 10 - dia, 11 - apenas, 12 - nunca, 13 - boa, 14 - agora, 15 - forma, 16 - mano, 17 - mundo, 18 - todo, 19 - qualquer, 20 - sabe, 21 - caso, 22 - todos, 23 - anos, 24 - site, 25 - dar, 26 - dinheiro, 27 - pois, 28 - onde, 29 - gente, 30 - ficar, 31 - obrigado, 32 - saber, 33 - menos, 34 - falar, 35 - realmente, 36 - vezes, 37 - quero, 38 - mim, 39 - talvez, 40 - verdade, 41 - forum, 42 - hoje, 43 - pouco, 44 - existe, 45 - conta, 46 - outros, 47 - outro, 48 - problema, 49 - f***, 50 - disso
\\ \hline
\end{tabular}
\end{table}

\begin{table}[!ht]
\caption{List of the 50 most frequent words in the relevant posts class from DATASETS II and III}
\label{tab_lista_100_palavras_mais_freq_rel}
\centering
\begin{tabular}{p{6.2cm}|p{6.2cm}}
\hline
\multicolumn{2}{c}{\textbf{Most Frequent Words}} \\
\hline
\multicolumn{2}{c}{\textbf{Relevant Posts}} \\
\hline
\textbf{DATASET II} & \textbf{DATASET III} \\ \hline
1 - dados, 2 - site, 3 - pessoa, 4 - conta, 5 - tipo, 6 - saber, 7 - pessoas, 8 - hacking, 9 - senha, 10 - link, 11 - hacker, 12 - linux, 13 - caso, 14 - melhor, 15 - onde, 16 - rede, 17 - nome, 18 - sabe, 19 - curso, 20 - faz, 21 - tempo, 22 - boa, 23 - qualquer, 24 - sites, 25 - informacoes, 26 - apenas, 27 - forum, 28 - mano, 29 - acesso, 30 - todos, 31 - sempre, 32 - tor, 33 - celular, 34 - forma, 35 - web, 36 - seguranca, 37 - aprender, 38 - agora, 39 - porque, 40 - pois, 41 - sistema, 42 - internet, 43 - google, 44 - email, 45 - dar, 46 - possivel, 47 - programacao, 48 - quero, 49 - virus, 50 - cpf
& 1 - dados, 2 - site, 3 - conta, 4 - pessoa, 5 - tipo, 6 - nome, 7 - boa, 8 - melhor, 9 - pessoas, 10 - onde, 11 - senha, 12 - telegram, 13 - faz, 14 - rede, 15 - informacoes, 16 - agora, 17 - saber, 18 - qualquer, 19 - cpf, 20 - acesso, 21 - forma, 22 - caso, 23 - porque, 24 - hacking, 25 - forum, 26 - dar, 27 - tempo, 28 - link, 29 - sites, 30 - mano, 31 - realmente, 32 - linux, 33 - sabe, 34 - obrigado, 35 - sistema, 36 - tor, 37 - email, 38 - sempre, 39 - possivel, 40 - apenas, 41 - facil, 42 - posso, 43 - seguranca, 44 - numero, 45 - outra, 46 - precisa, 47 - todos, 48 - celular, 49 - outro, 50 - dinheiro
\\ \hline
\end{tabular}
\end{table}

The analysis of the most frequent words in each dataset confirmed previous findings from the \emph{Relevant} posts identification test (Subsection \ref{subsec_testes_do_modelo_novos_posts}) and the LDA topic modeling analysis (Subsection \ref{subsec_resultados_lda}). For the new posts in \emph{DATASET III}, the model's classification aligned consistently with the labeled \emph{DATASET II}.

Several words are common across both datasets. For example, in the class of \emph{Not Relevant} posts (Table \ref{tab_lista_100_palavras_mais_freq_nao_rel}), words such as \emph{pessoas, vida, tempo}, and \emph{porque} appear at the top of the list in both datasets, distancing from a direct relation to cyber threats. In the class of \emph{Relevant} posts (Table \ref{tab_lista_100_palavras_mais_freq_rel}), words such as \emph{dados, site}, and \emph{conta} are at the top of the list, suggesting a possible data leak. Other words, such as \emph{senha} and \emph{CPF}, reinforce this hypothesis. Additionally, several other words shown in the graphs, such as \emph{hacker, hacking}, and \emph{vírus}, are directly related to cyber threats.


\subsection{Discussion and Comparison with Related Works} \label{subsec_comp_trabalhos_rel}

Our model for identifying malicious posts on the Dark Web using supervised machine learning has shown promising results. Building upon a carefully curated Portuguese-language dataset, we developed from scratch, it successfully identified relevant posts for obtaining and sharing CTI, achieving high confidence in distinguishing relevant from non-relevant content. The model's effectiveness was validated during evaluation with unlabeled data, where it detected potentially malicious content. These results confirm the robustness of the methodology applied, encompassing data collection, labeling, algorithm selection, and text representation techniques.

All related works cited in Section \ref{sec_background} share this article's goal of extracting relevant CTI information from unstructured data sources like the Dark Web and social networks. While many studies use different approaches, making direct comparisons challenging, three studies employed comparable methodologies. To better understand how our work compares to the state of the art, we decided to compare these three most closely related works according to the following attributes: Dataset, Vectorization, Algorithms, Objective, Evaluation, and Performance. Table \ref{tab_comparacao_com_trabalhos_rel} compares our work with these three studies.


\begin{sidewaystable}[!htp]
\centering
\resizebox{1.0\textwidth}{!}{ 
\begin{tabular}{|c|clccc|c|c|l|l|l|l|}
\hline
\multirow{2}{*}{\textbf{Work}} & \multicolumn{5}{c|}{\textbf{DATASET}}                                                                                                                                                                                                                                                                                                                                             & \multirow{2}{*}{\textbf{Vectorization}}                                                                       & \multirow{2}{*}{\textbf{Algorithms}}                                         & \multicolumn{1}{c|}{\multirow{2}{*}{\textbf{Objective}}}                                                         & \multicolumn{1}{c|}{\multirow{2}{*}{\textbf{Evaluation}}}                                                                               & \multicolumn{1}{c|}{\multirow{2}{*}{\textbf{Performance}}}                                                          & \multirow{2}{*}{\textbf{Observations}}                                                                                                                                                                                     \\ 
\hhline{~|-|-|-|-|-|} 
                                   & \multicolumn{1}{l|}{\textit{\textbf{Source}}}                                                    & \multicolumn{1}{l|}{\textit{\textbf{Labeling}}}                                                                                    & \multicolumn{1}{l|}{\textit{\textbf{Samples}}} & \multicolumn{1}{l|}{\textbf{Language}} & \multicolumn{1}{l|}{\textit{\textbf{Public}}}             &                                                                                                              &                                                                              & \multicolumn{1}{c|}{}                                                                                           & \multicolumn{1}{c|}{}                                                                                                                  & \multicolumn{1}{c|}{}                                                                                              &                                                                                                                                                                                                                           \\ \hline
\cite{dong2018new}                & \multicolumn{1}{c|}{Dark Web}                                                          & \multicolumn{1}{l|}{Manual}                                                                                                         & \multicolumn{1}{c|}{8,000}                      & \multicolumn{1}{c|}{EN}              & No                                                        & TF-IDF                                                                                                      & MLP                                                                          & \begin{tabular}[c]{@{}l@{}}Identify new threats\end{tabular}                                             & \begin{tabular}[c]{@{}l@{}}Used the threat \\intelligence platform \\(AlienVault OTX)\end{tabular}                              & 94\% accuracy                                                                                                   & High false positive rates                                                                                                                                                                                             \\ \hline
\cite{queiroz2019detecting}       & \multicolumn{1}{c|}{\begin{tabular}[c]{@{}c@{}}Dark Web / \\ Surface Web\end{tabular}} & \multicolumn{1}{l|}{Manual}                                                                                                         & \multicolumn{1}{c|}{9,470}                      & \multicolumn{1}{c|}{EN}              & \begin{tabular}[c]{@{}c@{}}URL does\\ not work\end{tabular} & \begin{tabular}[c]{@{}c@{}}Word2Vec/ Glove\\ Sent2vec / InferSent /\\ SentEncoder\end{tabular}              & SVM / CNN                                                                    & \begin{tabular}[c]{@{}l@{}}Enhance classification \\methods using \\embedding models \end{tabular} & \begin{tabular}[c]{@{}l@{}}Used as a basis a\\ previously developed work \end{tabular}                                      & \begin{tabular}[c]{@{}l@{}}96\% accuracy\\ 93\% recall\end{tabular}                                       & \begin{tabular}[c]{@{}l@{}}High false positive rates \\ causing low recall rates\end{tabular}                                                                                                             \\ \hline
\cite{koloveas2021intime}         & \multicolumn{1}{c|}{\begin{tabular}[c]{@{}c@{}}Dark Web /\\ Surface Web\end{tabular}}  & \multicolumn{1}{l|}{\begin{tabular}[c]{@{}l@{}}Occurrence of the\\ terms security and\\ IoT\end{tabular}}                             & \multicolumn{1}{c|}{1,677}                      & \multicolumn{1}{c|}{EN}              & No                                                        & Word2Vec                                                                                                    & SVM / RF                                                                     & \begin{tabular}[c]{@{}l@{}}Identify CTI \\information\end{tabular}                                      & There was no report                                                                                                                       & \begin{tabular}[c]{@{}l@{}}95\% accuracy\\ 61\% precision\\ 73\% recall\\ 64\% F-score\end{tabular} & \begin{tabular}[c]{@{}l@{}}Although the article mentions the use \\of data from the Dark Web and testing \\of various algorithms, the reported test\\ presented only Twitter data and the \\use of two algorithms: CNN and RF\end{tabular} \\ \hline
This work                   & \multicolumn{1}{c|}{Dark Web}                                                          & \multicolumn{1}{l|}{\begin{tabular}[c]{@{}l@{}}Occurrence of IoCs,\\ contextual keywords,\\ and manual analysis\end{tabular}} & \multicolumn{1}{c|}{26,575}                     & \multicolumn{1}{c|}{PT-BR}           & Yes                                                        & \begin{tabular}[c]{@{}c@{}}TF (Unigram e \\ Bigram) /\\ TF-IDF (Unigram e\\ Bigram) / Word2Vec\end{tabular} & \begin{tabular}[c]{@{}c@{}}SVM / RF / LR\\ LightGBM /\\ XGBoost\end{tabular} & \begin{tabular}[c]{@{}l@{}}Identification of \\relevant posts for CTI\end{tabular}                           & \begin{tabular}[c]{@{}l@{}}The model was evaluated\\ using new unlabeled data\\ with LDA topic modeling\end{tabular} & \begin{tabular}[c]{@{}l@{}}94\% accuracy\\ 74\% precision\\ 83\% recall\\ 79\% F-score\end{tabular}  & \begin{tabular}[c]{@{}l@{}}Accuracy rate for posts in the relevant\\ class above 83\%\end{tabular}                                                                                                                     \\ \hline
\end{tabular}%
}
\caption{Comparison with some related work}
\label{tab_comparacao_com_trabalhos_rel}
\end{sidewaystable}

Although the works cited in the table share similarities with this article, their objectives differ. Dong et al. \cite{dong2018new} focused on identifying new cyber threats on the Dark Web, while Koloveas et al. \cite{koloveas2021intime} concentrated on obtaining CTI information related to IoT devices. The objective of Queiroz et al. \cite{queiroz2019detecting} most closely aligns with this article's goal of identifying general CTI-relevant information in analyzed posts.

As shown in Table \ref{tab_comparacao_com_trabalhos_rel}, this article's dataset stands out due to its larger sample size, rigorous labeling criteria, and public availability. Regarding vectorization, while other studies used word frequency or word embedding techniques, this article evaluated both approaches. Queiroz et al. \cite{queiroz2019detecting} employed both word and sentence embedding techniques, aiming to compare their effectiveness against word frequency techniques from a previous study.

We tested more options regarding machine learning algorithms than the compared studies. While Koloveas et al. \cite{koloveas2021intime} mentioned using several algorithms, they presented results for only two. Our performance metrics were comparable to those of other studies, though some reported only limited metrics.

To evaluate the model for identifying content relevant to CTI, Dong et al. \cite{dong2018new}, which focused on identifying new threats, used an open threat platform as a baseline. In turn, Queiroz et al. \cite{queiroz2019detecting} compared results achieved using different vectorization techniques than those tested in a previous study. Meanwhile, Koloveas et al. \cite{koloveas2021intime} did not report any evaluation of their developed model. This article evaluated an unlabeled dataset involving LDA topic modeling and analysis of the most frequent words to compare the similarity between the labeled dataset and the dataset classified by the model.

\subsection{Implications for Practice}

The model proposed in this study has direct applicability in real-world CTI and Security Operations Center (SOC) environments. Its main contribution lies in its ability to classify posts from Dark Web forums as potentially malicious, which is particularly relevant in regions underrepresented in current CTI tools.

In practice, the model can be embedded into automated CTI pipelines, allowing for near real-time classification of newly crawled posts. Its probabilistic outputs provide a ranking mechanism that helps analysts prioritize alerts. This can reduce manual triage time and improve focus on high-relevance content. Furthermore, the model can be integrated into existing Threat Intelligence Platforms (TIPs), where it would serve as a filter or pre-processing stage for Dark Web content ingestion. By tagging posts as relevant or not, it can enrich contextual data, assist in the identification of IoCs, and contribute to proactive defense strategies.

Although the model was trained and validated using content written in Brazilian Portuguese, the underlying method is language-agnostic. The same multi-stage labeling process -- based on IoCs, contextual keywords, and manual validation -- can be adapted to forums in other languages with appropriate linguistic resources.

Importantly, the posts identified as relevant in this study reflect recurring threat patterns in underground communities. These include the sale of credentials, malware distribution, social engineering services, and instructions for financial fraud -- categories commonly linked to ransomware operations and APT campaigns~\cite{trendmicro2024midyear, socRadarbrazil2024}. These capabilities highlight the model’s potential to complement existing detection systems and enhance incident response efforts in organizations operating both locally and globally.


\section{Conclusion} \label{sec_conclusao}

This study aimed to explore Dark Web forums in search of new techniques and data sources for obtaining CTI, using text mining techniques and supervised machine learning. The goal was to develop a computational model capable of identifying relevant posts to assist the cybersecurity community in detecting threats, vulnerabilities, data leaks, and cyberattacks. From 26,575 posts collected from the \emph{Hidden Answers} and \emph{Deep Answers} forums, we generated two datasets: \emph{DATASET I}, where the labeling criterion was the simultaneous occurrence of IoCs and contextual keywords, and \emph{DATASET II}, which, in addition to the occurrence of IoCs and keywords, involved a manual analysis considering factors such as the category of the post. Five different text representations and five machine learning algorithms were tested, with the \emph{LightGBM and TF-IDF - Unigram} combination showing the best performance, achieving 94\% accuracy, 74\% precision, 79\% F-measure, and a recall rate of 83\%, representing the correct classification of 548 out of 658 samples analyzed.

Subsequently, 7,498 new posts were collected and classified (\emph{DATASET III}), confirming the accuracy achieved in the labeled dataset, with 15\% of the new posts identified as relevant. The LDA topic analysis revealed similarities between the topics of datasets II and III, demonstrating the model's ability to detect cyber threats in new posts. The analysis of the 100 most frequent words reinforced that the words in the relevant posts are directly related to threats or data leaks, demonstrating the robustness of the model developed in this study.

For future work, we suggest investigating the content of relevant posts to identify the different types of threats on the Dark Web; including other data sources beyond the Dark Web, such as Telegram channels; evaluating the model's performance on other data sources; testing other forms of text representation, such as BERT, to verify possible significant gains compared to the best representation we identified in this study (TF-IDF - Unigram); and finally, incorporating the generated model into an end-to-end system that collects, stores, analyzes, and generates alerts about malicious activities across different data sources.

\section*{Declarations}\label{declarations}

\subsubsection*{Funding}\label{funding}

The authors thank Datarisk for financial support.

\subsubsection*{Conflict of interest}\label{conflict_interest}

The authors have no competing interests or other interests that might be perceived to influence the results or discussion reported in this paper.

\subsubsection*{Ethics approval}\label{ethics_approval}

This manuscript adheres to the principles and policies of authorship ethics.

\subsubsection*{Consent to participate and publication}\label{consent_participate}

All authors read and approved the final manuscript for publication via the subscription publishing route.

\subsubsection*{Availability of data and materials}\label{availability_data_materials}

All materials used in this manuscript are public, and no permission is required. The results and data in this manuscript have not been published elsewhere.

\subsubsection*{Code availability}\label{code_availability}

All materials used in this manuscript are public, and no permission is required. Additional materials for this article are available at the following link: \hyperlink{https://github.com/sebastiaoafilho/Malicious_Posts_Identification}{Repository}

\subsubsection*{Authors' contributions}\label{authors_contributions}

The authors contributed equally to this work.


\bibliography{sn-bibliography}

\end{document}